\documentclass[letterpaper,12pt]{article}

\makeatletter
\DeclareFontFamily{OMX}{MnSymbolE}{}
\DeclareSymbolFont{MnLargeSymbols}{OMX}{MnSymbolE}{m}{n}
\SetSymbolFont{MnLargeSymbols}{bold}{OMX}{MnSymbolE}{b}{n}
\DeclareFontShape{OMX}{MnSymbolE}{m}{n}{
    <-6>  MnSymbolE5
   <6-7>  MnSymbolE6
   <7-8>  MnSymbolE7
   <8-9>  MnSymbolE8
   <9-10> MnSymbolE9
  <10-12> MnSymbolE10
  <12->   MnSymbolE12
}{}
\DeclareFontShape{OMX}{MnSymbolE}{b}{n}{
    <-6>  MnSymbolE-Bold5
   <6-7>  MnSymbolE-Bold6
   <7-8>  MnSymbolE-Bold7
   <8-9>  MnSymbolE-Bold8
   <9-10> MnSymbolE-Bold9
  <10-12> MnSymbolE-Bold10
  <12->   MnSymbolE-Bold12
}{}

\let\llangle\@undefined
\let\rrangle\@undefined
\DeclareMathDelimiter{\llangle}{\mathopen}%
                     {MnLargeSymbols}{'164}{MnLargeSymbols}{'164}
\DeclareMathDelimiter{\rrangle}{\mathclose}%
                     {MnLargeSymbols}{'171}{MnLargeSymbols}{'171}
\makeatother

\usepackage{array}
\usepackage{jheppub}

\DeclareMathOperator{\Conf}{Conf}
\DeclareMathOperator{\Gr}{Gr}

\newcommand{\hel}{\textrm{k}}
\newcommand{\nkmhv}[1]{\textrm{N}^{#1}\textrm{MHV}}

\newcommand{\eqnRef}[1]{Eq.~(\ref{eqn:#1})}
\newcommand{\figRef}[1]{Fig.~\ref{fig:#1}}

\newcommand{\secRef}[1]{Sec.~\ref{sec:#1}}
\newcommand{\tabRef}[1]{Tab.~\ref{tab:#1}}

\newcommand{\ko}[1]{\mathcal{K}_{#1}}
\newcommand{\ro}[1]{\mathcal{R}_{#1}}
\newcommand{\uo}[1]{\mathcal{U}_{#1}}

\newcommand{\nEmpty}{n_{\textrm{empty}}}
\newcommand{\nFull}{n_{\textrm{filled}}}

\newcommand{\ntri}{n_{\textrm{tri}}}
\newcommand{\nhigh}{n_{\textrm{high}}}

\title{Boundaries of Amplituhedra and
NMHV Symbol Alphabets at Two Loops}

\author{I.~Prlina,$^1$}
\author{M.~Spradlin,$^{1,2}$}
\author{J.~Stankowicz$^{1,3}$}
\author{and S.~Stanojevic$^1$}

\affiliation{$^1$ Department of Physics,
  Brown University,
  Providence RI 02912}

\affiliation{$^2$ School of Natural Sciences,
  Institute for Advanced Study,
  Princeton NJ 08540}

\affiliation{$^3$ Kavli Institute for Theoretical Physics,
  University of California,
  Santa Barbara CA 93106}

\emailAdd{igor\_prlina@brown.edu}
\emailAdd{marcus\_spradlin@brown.edu}
\emailAdd{james\_stankowicz@brown.edu}
\emailAdd{stefan\_stanojevic@brown.edu}

\abstract{
In this sequel to arXiv:1711.11507 we classify the boundaries
of amplituhedra relevant for determining
the branch points of general two-loop amplitudes in
planar $\mathcal{N}=4$ super-Yang--Mills theory.
We explain the connection to on-shell diagrams, which serves
as a useful cross-check.
We determine the branch points of all two-loop NMHV amplitudes
by solving the Landau equations for the relevant configurations
and are led thereby to a conjecture for the symbol alphabets
of all such amplitudes.
}

\begin{document}
\maketitle

\section{Introduction}

It has been a long-standing goal to determine
scattering amplitudes in quantum field theory from knowledge
of their analytic structure coupled with other basic physical
and mathematical input.
In planar $\mathcal{N}=4$ super-Yang--Mills theory (which
we refer to as SYM theory), the current state of the art for
carrying out explicit computations of multi-loop amplitudes
is a bootstrap program that relies fundamentally
on assumptions about the location of branch points of certain
amplitudes.

The aim of the research program initiated
in~\cite{Dennen:2015bet,Dennen:2016mdk} for MHV amplitudes
and generalized to non-MHV amplitudes in~\cite{Prlina:2017azl}
(to which this paper should be considered a sequel)
is to provide an \emph{a priori} derivation
of the set of branch points for any given amplitude.
For sufficiently simple amplitudes in SYM theory\footnote{General
amplitudes lie outside the class of generalized
polylogarithm functions that have well-defined
symbols, see for
example~\cite{CaronHuot:2012ab,Nandan:2013ip,Bourjaily:2017bsb}
for a discussion of this in the context of SYM theory.}
this information can go a long way by leading to natural
guesses for the \emph{symbol alphabets}~\cite{Goncharov:2010jf}
of various amplitudes.
The possibility to do so exists because of the simple fact
pointed out in~\cite{Maldacena:2015iua}
that the locus in the space of external data
$\Conf_n(\mathbb{P}^3)$ where
the symbol letters of a given amplitude vanish
should be the same as the locus where the corresponding
Landau equations~\cite{Landau:1959fi,ELOP}
admit solutions.  A slight refinement of this statement,
to account for the fact that amplitudes in general have algebraic
branch cuts in addition to logarithmic cuts, was discussed
in Sec.~7 of~\cite{Prlina:2017azl}.

The hexagon bootstrap program, which
has succeeded in computing
all six-point amplitudes through
five loops~\cite{Dixon:2011pw,Dixon:2011nj,Dixon:2013eka,Dixon:2015iva,Caron-Huot:2016owq}, relies on the hypothesis
that these amplitudes can have branch points only at
nine specific loci in the space of external data
$\Conf_6(\mathbb{P}^3)$.  Similarly the heptagon
bootstrap~\cite{Drummond:2014ffa},
which has revealed
the symbols of the seven-point four-loop
MHV and three-loop NMHV amplitudes~\cite{Dixon:2016nkn},
assumes 42 particular branch points.
Ultimately we may hope for an all-loop proof of these hypotheses
about six- and seven-point amplitudes, but in this paper
we focus on the less ambitious goal of deriving the singularity
loci for all two-loop NMHV amplitudes in SYM theory.
The result, summarized in~\secRef{symbol-alphabets}, leads to
a natural conjecture for the symbol alphabets of these amplitudes
which we hope may be employed in the near future by bootstrappers
eager to study this class of amplitudes.

The rest of this paper is organized as follows.
In~\secRef{classification}
we develop a procedure for constructing certain boundaries
of two-loop amplituhedra by ``merging'' one-loop configurations
of the type classified in the prequel~\cite{Prlina:2017azl}.
In~\secRef{presentation} we organize the results according to
helicity and codimensionality (the number of on-shell conditions
satisfied by each configuration) and discuss some subtleties
about overconstrained configurations that
require resolution.
Section~\ref{sec:on-shell-diagrams} discusses the connection between
branches of solutions to on-shell conditions and on-shell diagrams,
which provides a useful cross-check of our classification.
In~\secRef{nmhv-landau-analysis} we discuss the analysis of
the Landau equations for configurations relevant for NMHV
amplitudes and, in Eqns.~(\ref{eqn:twoloopalphabet})
and~(\ref{eqn:nmhvsymbolalphabets}), we present a conjecture for
the symbol alphabets of all two-loop NMHV amplitudes.

\section{Classification of Two-Loop Boundaries}
\label{sec:classification}

In this section we classify certain boundaries of two-loop amplituhedra.
This analysis builds heavily on Sections~3--5
of~\cite{Prlina:2017azl}, and in particular we show how to recycle
the one-loop boundaries classified there by ``merging'' pairs of
one-loop boundaries into two-loop boundaries.
We find that two different formulations
of the amplituhedron --- the original
formulation in terms of $C$ and $D$
matrices~\cite{Arkani-Hamed:2013jha},
and the reformulation in terms of sign flips~\cite{Arkani-Hamed:2017vfh} ---
play two complementary roles, exactly as in~\cite{Prlina:2017azl}.
Specifically, the former is useful for establishing the existence
of boundaries by constructing explicit $C$ and $D$ matrix representatives,
while the latter is useful for establishing the non-existence of any
other boundaries.

Before proceeding let us dispense of some important details that
would otherwise overcomplicate our exposition.
There is a parity symmetry between $A_{n,\hel,L}$,
the $n$-point, $\nkmhv{k}$, $L$-loop amplitude in SYM theory,
and its parity conjugate $A_{n,n-\hel-4,L}$.
For fixed $n$, amplitudes become increasingly complicated as $\hel$ is increased
from zero, but after $\hel \sim n/2$ they must begin to decrease
in complexity until the upper bound $\hel = n-4$.
In what follows we will often make use of lower bounds on $\hel$,
or on constructions that increment $\hel$ by 1.
In making these arguments, we always have in mind
that $\hel$ is sufficiently small compared to $n$.
In other words, unless otherwise stated,
we are always working in the ``low-$\hel$'' regime, to use
the terminology of~\cite{Prlina:2017azl}.  At the very end of our
analysis, once we have all of the desired results in this
regime, we appeal to parity symmetry in order
to translate low-$\hel$ results into high-$\hel$ results.
However the details of matching these two regimes
near the midpoint $\hel \sim n/2$
can be quite intricate, even moreso at two loops than it
was in the one-loop analysis of~\cite{Prlina:2017azl}.

\subsection{Identifying the Relevant Boundaries}
\label{sec:identifying}

In general,
a configuration $(Y, \mathcal{L}^{(1)}, \mathcal{L}^{(2)})$ lies on a
boundary of a two-loop amplituhedron if
at least one item on the following menu is satisfied:
\begin{enumerate}
\renewcommand*\labelenumi{(\theenumi)}
\item $Y$ is such that some
four-brackets of the form $\langle a\,a{+}1\,b\,b{+}1\rangle$
vanish,
\item
$\mathcal{L}^{(1)}$ satisfies some
on-shell conditions
$\langle \mathcal{L}^{(1)}\,a_1\,a_1{+}1\rangle = \cdots =
\langle \mathcal{L}^{(1)}\,a_{d_1}\,a_{d_1}{+}1\rangle = 0$,
\item
$\mathcal{L}^{(2)}$ satisfies some
on-shell conditions
$\langle \mathcal{L}^{(2)}\,b_1\,b_1{+}1\rangle = \cdots =
\langle \mathcal{L}^{(2)}\,b_{d_2}\,b_{d_2}{+}1\rangle = 0$,
\item
or $\langle \mathcal{L}^{(1)}\,\mathcal{L}^{(2)}\rangle = 0$.
\end{enumerate}
Above and through the remainder of the paper, we always take
$\langle ABCD \rangle \equiv [Y ABCD]$ --- what we call projected
four-brackets following~\cite{Arkani-Hamed:2017vfh}.

For the purpose of finding Landau singularities
we are always interested only in loop momenta $(\mathcal{L}^{(1)},
\mathcal{L}^{(2)})$ that exist for generic projected external
data, i.e., for generic $Y$, so we disregard
possibility (1) in all that follows.  Next, we note that for configurations
which do not satisfy (4),
the Landau equations decouple into two separate sets
of equations on the two individual loop momenta, so there
can be no new Landau singularities beyond those already found
at one loop.
Therefore in all that follows we only consider boundaries on which
$\langle \mathcal{L}^{(1)}\,\mathcal{L}^{(2)}\rangle = 0$.
The Landau equations similarly degenerate if either $d_1$ or
$d_2$ (defined in the preceeding paragraph) is zero, so we are only interested in configurations
with $d_1 d_2 > 0$.

The above considerations motivate us to define
an $\mathcal{L}$-\emph{boundary} of a two-loop amplituhedron
as a configuration $(Y, \mathcal{L}^{(1)}, \mathcal{L}^{(2)})$ for which
$Y$ is such that the projected external data are generic,
$\langle \mathcal{L}^{(1)}\,\mathcal{L}^{(2)}\rangle = 0$, and
each $\mathcal{L}$ satisfies at least one on-shell condition
of the form $\langle \mathcal{L}\,a\,a{+}1\rangle = 0$.
In particular, these conditions imply that
both $(Y, \mathcal{L}^{(1)})$ and $(Y,\mathcal{L}^{(2)})$ must lie
on boundaries
of some one-loop amplituhedra; each of these must therefore
be one of the 19 branches tabulated
in Tab.~1 of~\cite{Prlina:2017azl}.

\subsection{Merging One-Loop Boundaries}
\label{sec:merging}

The preceding analysis suggests that
the boundaries of two-loop amplituhedra can be
understood by merging various one-loop boundaries.
Let us now see how this works in detail.
Suppose that $(Y^{(1)}, \mathcal{L}^{(1)})$ and $(Y^{(2)}, \mathcal{L}^{(2)})$
lie on boundaries of $\mathcal{A}_{n,\hel_1,1}$
and $\mathcal{A}_{n,\hel_2,1}$, respectively.
Then they can be represented as
$Y^{(\alpha)} = C^{(\alpha)} \mathcal{Z}$ and
$\mathcal{L}^{(\alpha)} = D^{(\alpha)} \mathcal{Z}$,
where for each $\alpha \in \{1, 2\}$, the matrices
$C^{(\alpha)}$,
$\left( \begin{smallmatrix} D^{(\alpha)} \\ C^{(\alpha)}
\end{smallmatrix} \right)$,
and $D^{(\alpha)}$ (as shown in~\cite{Prlina:2017azl}), are
all non-negative.
In order to streamline the argument we initially consider
$\hel_1$ and $\hel_2$ to be the smallest values of helicity for which
boundaries of the desired class exist, and we take each pair
$(C^{(\alpha)}, D^{(\alpha)})$ to have the form of one of the 19 branches
shown in Secs.~4.2 through 4.4 of~\cite{Prlina:2017azl}.
We will show that such a pair of valid one-loop boundary
configurations can be uplifted into a valid two-loop
boundary configuration $(C, D^{(1)}, D^{(2)})$ satisfying
$\langle \mathcal{L}^{(1)} \, \mathcal{L}^{(2)} \rangle = 0$ by
constructing an appropriate matrix $C$ from $C^{(1)}$ and $C^{(2)}$.

The process of merging two boundaries depends
on whether the two loop momenta $\mathcal{L}^{(1)}$,
$\mathcal{L}^{(2)}$ each pass through some
common external point $Z_i$. If they do, then
we say that they \emph{manifestly intersect} and the
condition that $\langle \mathcal{L}^{(1)}\,\mathcal{L}^{(2)}\rangle = 0$
is automatically satisfied.
In this case we can simply stack the two individual $C$-matrices on top of each
other in order to form
\begin{align}
C = \left( \begin{matrix}
C^{(1)} \\
C^{(2)} \end{matrix}\right).
\label{eqn:mergedC}
\end{align}
If, on the other hand, the two loop momenta do not manifestly
intersect, then we can still ensure that
$\langle \mathcal{L}^{(1)}\,\mathcal{L}^{(2)}\rangle = [ (C \mathcal{Z})\,
\mathcal{L}^{(1)} \, \mathcal{L}^{(2)}] = 0$ by adding
one additional suitably crafted row to $C$.  Specifically,
if $A^{(\alpha)}$, $B^{(\alpha)}$ are any
four points in $\mathbb{P}^n$
such that $\mathcal{L}^{(\alpha)} = (A^{(\alpha)} \mathcal{Z},
B^{(\alpha)} \mathcal{Z})$,
then adding a row to $C$ that is any linear combination of
these four points will guarantee
that $\langle \mathcal{L}^{(1)}\,\mathcal{L}^{(2)}\rangle = 0$.

In this manner we have constructed a candidate
for a configuration on the boundary of $\mathcal{A}_{n,\hel,2}$
with $\hel = \hel_1 + \hel_2$ in the case of manifest intersection,
or $\hel = \hel_1 + \hel_2 + 1$ otherwise.
It remains to verify that this configuration is \emph{valid},
which means that $C$ can be chosen so that it and the matrices
$\left( \begin{smallmatrix} D^{(1)} \\ C\end{smallmatrix} \right)$,
$\left( \begin{smallmatrix} D^{(2)}\\ C \end{smallmatrix} \right)$,
and
$
\left( \begin{smallmatrix}
D^{(1)}
\\
D^{(2)}
\\
C
\end{smallmatrix}\right)
$
are all non-negative.

\subsection{Planarity from Positivity}

Let us begin by analyzing the non-negativity of the
$C$-matrix shown in~\eqnRef{mergedC}.
The nonzero columns of each $C^{(\alpha)}$ (which may be read off
from Secs.~4.3 and 4.4 of~\cite{Prlina:2017azl})
are grouped into clusters corresponding to the sets of contiguous indices
appearing in the on-shell conditions satisfied by the corresponding
$\mathcal{L}^{(\alpha)}$.  For example,
for a boundary on which the three-mass
triangle on-shell conditions $\langle \mathcal{L}\, i\, i{+}1\rangle
= \langle \mathcal{L}\, j\, j{+}1\rangle =
\langle \mathcal{L}\, k\, k{+}1\rangle = 0$ are satisfied,
the $C$-matrix is zero
except in six columns grouped into three clusters
$\{i, i{+}1\}$, $\{j, j{+}1\}$ and $\{k, k{+}1\}$.

When we stack two $C$-matrices together, the result can be one of two different
cases depending on whether or not the clusters of $C^{(1)}$
are cyclically adjacent compared to the clusters of $C^{(2)}$.
If so, then the stacked $C$-matrix has the schematic form
\begin{equation}
C = \left( \begin{matrix}
C^{(1)} \\
C^{(2)} \end{matrix}\right)=
\left(
\begin{array}{*{11}c}
\cdots & 0 & \star & 0 & \star & 0 & 0 & 0 & 0 & 0 & \cdots
\cr
\cdots & 0 & 0 & 0 & 0 & 0 & \star & 0 & \star & 0 & \cdots
\end{array}
\right) \
\begin{matrix}
\} \ \hel_1 \ \textrm{rows} \\
\} \ \hel_2 \ \textrm{rows}
\end{matrix}
\label{eqn:planar}
\end{equation}
which we call \emph{planar}; otherwise it is of the form
\begin{align}
C = \left( \begin{matrix}
C^{(1)} \\
C^{(2)} \end{matrix}\right)=
\left( \begin{array}{*{11}c}
\cdots & 0 & \star & 0 & 0 & 0 & \star & 0 & 0 & 0 & \cdots
\cr
\cdots & 0 & 0 & 0 & \star & 0 & 0 & 0 & \star & 0 & \cdots
\end{array}
\right)
\begin{matrix}
\} \ \hel_1 \ \textrm{rows\,\hphantom{.}} \\
\} \ \hel_2 \ \textrm{rows\,.}
\end{matrix}
\label{eqn:nonplanar}
\end{align}
which we call \emph{non-planar}.
In Eqns.~(\ref{eqn:planar}) and~(\ref{eqn:nonplanar}) each $\star$ is
shorthand for one or more contiguous columns (i.e,
clusters) of non-zero entries,
and we suppress displaying columns shared by the two $C$-matrices,
which are not relevant to our argument.
Also as indicated the top (bottom) row is shorthand for
$\hel_1$ ($\hel_2$) rows.
Given that our starting point is a pair of matrices
$C^{(1)}$, $C^{(2)}$ that are each non-negative, it is clear that the resulting
stacked $C$-matrix
has a chance to be non-negative (for certain values of its parameters) only for planar configurations;
the minors of~\eqnRef{nonplanar} manifestly have non-definite signs.

In cases when $\mathcal{L}^{(1)}$ and $\mathcal{L}^{(2)}$ do not
manifestly intersect we need to add an additional row to $C$
as described in the previous section.
This additional row can be considered part of either $C^{(1)}$ or $C^{(2)}$.
Since the coefficients in this row can be arbitrary and still
preserve $\langle \mathcal{L}^{(1)}\,\mathcal{L}^{(2)}\rangle=0$,
the coefficients can always be chosen such that the enlarged
$C$-matrix is non-negative.
The conclusion that only planar $C$'s can be made positive
still holds.

The nomenclature of `planar' and `non-planar' clusters is appropriate
in light of the fact that the locations of the clusters precisely correspond
to the sets of indices appearing in on-shell conditions listed in points (2)
and (3) at the beginning of \secRef{identifying}.
In a configuration like~\eqnRef{planar} there exist $a$, $b$
such that all of the on-shell conditions satisfied by
$\mathcal{L}^{(1)}$ lie in the range $\{a, a{+}1, \ldots, b, b+1\}$
while all of the on-shell conditions satisfied by $\mathcal{L}^{(2)}$
lie in the range $\{b, b{+}1, \ldots, a, a{+}1\}$ (as usual, all indices
are always understood mod $n$). Consequently, the two-loop
Landau diagram depicting the merged sets of on-shell
conditions (together with the propagator
$\langle \mathcal{L}^{(1)}\,\mathcal{L}^{(2)}\rangle$
shared between the two loops) is planar.
By the same argument, a nonplanar configuration
such as~\eqnRef{nonplanar}
is necessarily associated to
a nonplanar Landau diagram.

Now let us consider the non-negative matrices
$\left( \begin{smallmatrix} D^{(\alpha)} \\ C^{(\alpha)}\end{smallmatrix} \right)$
for the two individual initial boundary configurations ($\alpha=1$ or $2$).
We require that
these matrices stay non-negative when $C^{(\alpha)}$ is replaced by $C$.
By the argument given in Sec.~4.7 of~\cite{Prlina:2017azl},
this will be the case if the rows added to $C^{(\alpha)}$ have
nonzero entries only in the gaps between clusters of $C^{(\alpha)}$.
But this is just another way to phrase the planarity condition described
above, so again we see that planarity is enforced, this
time by requiring non-negativity of
$\left( \begin{smallmatrix} D^{(\alpha)} \\ C\end{smallmatrix} \right)$.

The final step in establishing the validity
of the configuration $(C, D^{(1)}, D^{(2)})$
is checking that
the matrix
$\left( \begin{smallmatrix} D^{(1)} \\
D^{(2)} \\
C\end{smallmatrix} \right)$
is non-negative.
In the parameterization we have chosen, all of the maximal minors
of this matrix actually vanish.  If the two loops manifestly
intersect this can be checked by looking
at the form of the $(C, D)$ matrices tabulated
in~\cite{Prlina:2017azl}.  If they do not manifestly intersect the analysis
is even easier, since in such cases we have included in $C$ a row
that is some linear combination of the four rows of $D^{(1)}, D^{(2)}$.

The argument as presented appears to fail
if either of the individual one-loop
boundaries is MHV, in which case there is no $C$ matrix.
However, for MHV boundaries it can
be seen from the expressions tabulated in
Sec.~4.2 of~\cite{Prlina:2017azl} that the $D$-matrix serves
the same role as the $C$-matrix played in the above argument.
For example, if $\hel_1 = 0$ so that $C^{(1)}$ is empty,
then $C = C^{(2)}$
so the requirement that
$\left( \begin{smallmatrix} D^{(1)} \\ C\end{smallmatrix} \right) =
\left( \begin{smallmatrix} D^{(1)} \\ C^{(2)}\end{smallmatrix} \right)$
must be non-negative requires that the clusters of $D^{(1)}$ be cyclically
adjacent compared to the clusters of $C^{(2)}$.
If both $\hel_1$ and $\hel_2$ are zero then $C$ is empty
and the same conclusion
follows from consideration
of the matrix
$
\left( \begin{smallmatrix}
D^{(1)}
\\
D^{(2)}
\end{smallmatrix}\right)
$.
Therefore, in all cases, the various non-negativity conditions
imply that the Landau diagram must be planar.
This emergent planarity was discussed
in context of MHV amplitudes in~\cite{Arkani-Hamed:2013kca}.

In conclusion, we have established that a boundary of
$\mathcal{A}_{n,\hel,2}$ can be constructed by ``merging'' a
boundary of $\mathcal{A}_{n,\hel_1,1}$
with a boundary of $\mathcal{A}_{n,\hel_2,1}$,
with $\hel - \hel_1 - \hel_2 = 0$ or $1$ depending
on whether $\mathcal{L}^{(1)}$ and $\mathcal{L}^{(2)}$ manifestly intersect.
So far we have considered $\hel_1$ and $\hel_2$ to saturate the
lower bounds shown in Tab.~1 of~\cite{Prlina:2017azl}, but once a valid
configuration $(C, D^{(1)}, D^{(2)})$ has been constructed as described in this
section, it can be lifted to higher values of $\hel$
by growing the $C$-matrix according to a suitably modified version
of the argument given in Sec.~4.7 of that reference.

\subsection{Establishing the Lower Bound on Helicity}

We have shown that it is possible to merge two one-loop boundaries
with (minimal) helicities $\hel_1$ and $\hel_2$ in order to generate
two-loop boundaries with helicities
$\hel \ge \hel_1 + \hel_2$.
The merging algorithm we have described cannot generate
boundaries with $\hel$ below this lower bound.
In this section we prove that we have not overlooked any
potential two-loop boundaries. To do so, we
use the formulation of
amplituhedra in terms of sign flips~\cite{Arkani-Hamed:2017vfh}
(reviewed also in Sec.~2.2 of~\cite{Prlina:2017azl}) in order
to prove the lower bound.

The proof is essentially a loop-level version of the factorization
argument presented in Sec.~6 of~\cite{Arkani-Hamed:2017vfh} for
tree-level amplituhedra.
Let $(\mathcal{L}^{(1)}, \mathcal{L}^{(2)})$ be some configuration
of loop momenta on some codimension $d_1 + d_2 + 1$ boundary
of $\mathcal{A}_{n,\hel,2}$, satisfying the on-shell
conditions
\begin{align}
\label{eqn:l1cuts}
\langle \mathcal{L}^{(1)}\,a_1\,a_1{+}1\rangle = \cdots
= \langle \mathcal{L}^{(1)}\,a_{d_1}\,a_{d_1}{+}1\rangle &= 0\,, \\
\langle \mathcal{L}^{(2)}\,b_1\,b_1{+}1\rangle = \cdots
= \langle \mathcal{L}^{(2)}\,b_{d_2}\,b_{d_2}{+}1\rangle  =
\langle \mathcal{L}^{(1)}\,\mathcal{L}^{(2)}\rangle&= 0\,,
\label{eqn:l2cuts}
\end{align}
with the sets of indices $\{ a_1, \ldots, a_{d_1}\}$ and $\{b_1,\ldots, b_{d_2}\}$
cyclically ordered and
with $1 \le d_1, d_2 \le 4$ as detailed in~\cite{Prlina:2017azl}.
Planarity requires that all of the $b$'s fall inside
an interval between two consecutive $a$'s; specifically,
there exists some $j$ such that $a_j \le b_i \le a_{j+1}$
for all $i$.  Once we have identified this value of $j$, let's backtrack
and consider factorization
(as described in~\cite{Arkani-Hamed:2017vfh})
on the boundary
$\langle \mathcal{L}^{(1)}\,a_j\,a_j{+}1\rangle =
\langle \mathcal{L}^{(1)}\,a_{j+1}\,a_{j+1}{+}1\rangle = 0$.
Then $\mathcal{L}^{(1)}$ passes through some point $A$ on the line
$(a_j\, a_{j}{+}1)$ and some point $B$ on the line
$(a_{j+1}\, a_{j+1}{+}1)$.  With $\mathcal{L}^{(1)} = (A\,B)$
we consider the sets of momentum twistors
\begin{align}
V &= \{ A, Z_{a_j+1}, \ldots, Z_{a_{j+1}}, B\}\,,\\
W &= \{ B, Z_{a_{j+1}+1},\ldots, Z_{a_j}, A\}\,.
\end{align}
Thinking of $V$ and $W$ separately as ``(projected) external data''
for sub-amplituhedra describing
two smaller sets of scattering
particles\footnote{We put ``(projected) external data'' in quotation marks when it is (projected)
external data only for a sub-amplituhedron, not for the full amplituhedron.},
it follows using arguments analogous to those
in Sec.~6 of~\cite{Arkani-Hamed:2017vfh} that they lie in the principal
domain for helicities $\hel_V$ and $\hel_W$ satisfying
$\hel_V + \hel_W = \hel$ where $\hel$ is the original helicity
sector of the (projected) external data $\{Z_i\}$.

Under the assumption that the two-loop configuration
$(Y,\mathcal{L}^{(1)},\mathcal{L}^{(2)})$ is
a boundary of $\mathcal{A}_{n,\hel,2}(Z)$, we prove below the following statements:
\begin{itemize}
\item if $\mathcal{L}^{(1)}$ is a solution to
the on-shell conditions~(\ref{eqn:l1cuts})
with minimum helicity $\hel_1$,
then $\hel_V \geq \hel_1$,
and similarly,
\item if $\mathcal{L}^{(2)}$ is a solution to
the on-shell conditions~(\ref{eqn:l2cuts})
with minimum helicity $\hel_2$,
then $\hel_W \geq \hel_2$.
\end{itemize}
Once we show this, it follows immediately that
the two-loop configuration $(\mathcal{L}^{(1)},
\mathcal{L}^{(2)})$ cannot be a valid boundary unless
\begin{equation}
\label{eqn:k-inequality}
\hel = \hel_V + \hel_W \geq \hel_1 + \hel_2\,.
\end{equation}

\paragraph{Proof.}
The minimum values of helicity $\hel_{\rm min}$ for which sets of
one-loop one-shell conditions admit solutions inside
the closure of $\mathcal{A}_{n,\hel,1}$ were
derived in Sec.~4 of~\cite{Prlina:2017azl}.
In that analysis, the fact that a set of on-shell conditions does not have
valid solutions of a certain type for $\hel < \hel_{\rm min}$ followed
from the fact that the non-negativity constraints on the $C$ and
$\left( \begin{smallmatrix} D \\ C \end{smallmatrix} \right)$
matrices required certain sequences of (projected) four-brackets
to contain at least $\hel_{\rm min}$ sign flips.
In analyzing the constraints on the solution $\mathcal{L}^{(1)}$
to~\eqnRef{l1cuts}, the relevant sequences of four-brackets
are of the form $\langle \alpha\, \beta\, \gamma\, \bullet \rangle$
where $\alpha$, $\beta$ and $\gamma$ are functions
of the momentum twistors belonging
to the set $S = \{ Z_{a_1}, Z_{a_1+1}, \cdots, Z_{a_{d_1}},
Z_{a_{d_1}+1}\}$ only, and the required sign flips occur
between adjacent entries in $S$.
Note that there are two points ($Z_{a_j}$ and
$Z_{a_{j+1}{+}1}$) in $S$ that lie outside $V$,
the ``(projected) external data'' for one of the sub-amplituhedra
under consideration.
However, because $A$ lies on the line $(a_j\,a_j{+}1)$
and $B$ lies on the line $(a_{j+1}\,a_{j+1}{+}1)$, we clearly
have $(a_j\, a_j{+}1) = (a_j\,A)$ and
similarly $(a_{j+1}\,a_{j+1}{+}1) =
(a_{j+1}\,B)$ so we can choose to express $\alpha$, $\beta$ and $\gamma$
in terms of momentum twistors belonging to
\begin{align}
S' = \{ Z_{a_1}, Z_{a_1+1}, \ldots
Z_{a_j}, A, B, Z_{a_{j+1}+1}, \ldots, Z_{a_{d_1}},
Z_{a_{d_1}+1} \} \subset V\,.
\end{align}
Therefore the abovementioned sequences can all be expressed
in terms of the ``(projected) external data'' associated to the $V$
sub-amplituhedron. Since there are $\hel_1$ sign flips in $S'$, it
must be the case that $\hel_V \ge \hel_1$.
It follows similarly that $\hel_W \ge \hel_2$. $\blacksquare$

In~\eqnRef{k-inequality} we derived an inequality
$\hel \ge \hel_1+\hel_2$, and at the end of~\secRef{merging}
we explained that two-loop configurations have support
starting from $\hel = \hel_1 + \hel_2$ or
$\hel = \hel_1 + \hel_2+1$.
In~\secRef{merging}
we effectively defined $\hel_1$ and $\hel_2$
as the minimum helicities for configurations of loop momenta
satisfying sets of disjoint on-shell conditions, not
including the shared propagator.
However, in this section the definition of $\hel_2$ (only) now
includes the shared propagator (c.f.~\eqnRef{l2cuts}).
Effectively, this means that
the $\hel_2$ here is the same as in~\secRef{merging} only
for manifest intersection, but one greater than the latter in the
case of non-manifest intersection.

\section{Presentation of the Results}
\label{sec:presentation}

It is now a straightforward exercise to explicitly enumerate
all possible pairs of one-loop boundaries, using those listed
in Tab.~1 of~\cite{Prlina:2017azl}, and to determine the minimum
value of $\hel$ such that the merged configuration is a valid
boundary of $\mathcal{A}_{n,\hel,2}$.
The resulting set is too large to display in a single figure
of the type of Fig.~1 of~\cite{Prlina:2017azl} (which is a summary
of the analogous results at one loop), so we focus first on
the maximal codimension boundaries. Each involves
a total of $d=8$ on-shell conditions:
the shared condition $\langle \mathcal{L}^{(1)}\, \mathcal{L}^{(2)}\rangle = 0$
together with seven conditions on the two loop momenta
($d_1 + d_2 = 7$, in the notation of~\secRef{identifying}).

%
%
%
\begin{figure}
\centering
\includegraphics[width=5.6in]{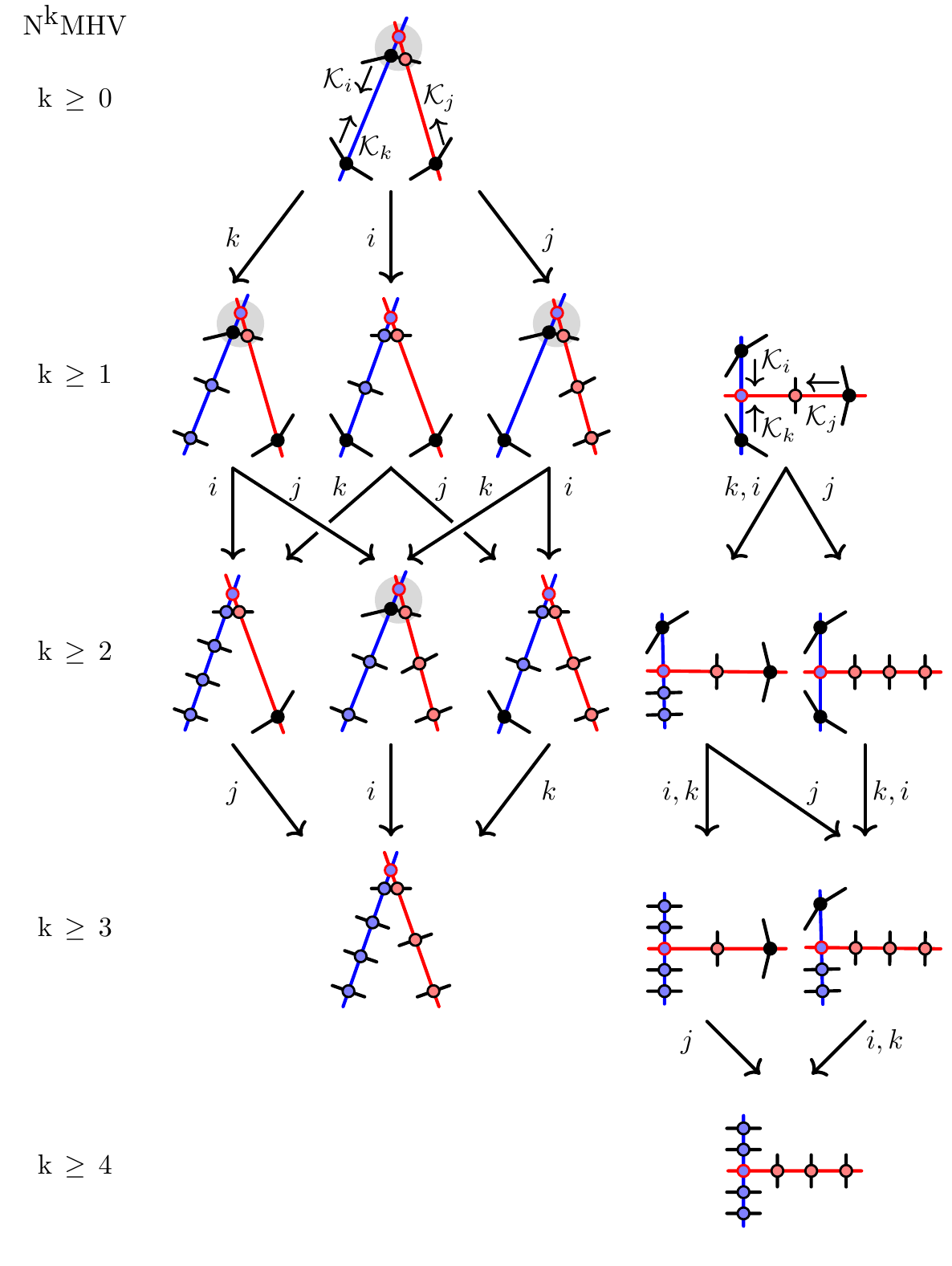}
\caption[Graph flow.]{%
The twistor diagrams depicting the 14 distinct
maximal codimension boundaries of two-loop $\nkmhv{\hel}$ amplituhedra.
See the text for more details.
}
 \label{fig:seed-amplituhedron-diagrams}
\end{figure}
%
%
%

We find a total of 14 topologically distinct maximal codimension configurations at two loops, which are summarized
in~\figRef{seed-amplituhedron-diagrams}.  The figure emphasizes
the fact that all 14 varieties of
$\mathcal{L}$-boundaries can be obtained by some sequence of
helicity-increasing
operations $\ko{}$ (defined in Sec.~5.2
of~\cite{Prlina:2017azl}) acting on just two primitive diagrams,
one at MHV level and one at NMHV level.
The entirety of this figure should be thought of as the two-loop
$d=8$ analog of the one-loop $d=4$ column of Fig.~2 of that reference.
In the figure, an arrow labeled by $i$ indicates that the diagram
at the end of the arrow can be obtained by acting with
$\ko{i}$ on the diagram at the beginning of the arrow.
An arrow carries two labels if the result of acting with two
different instances of $\ko{}$ gives topologically equivalent diagrams,
in which case only the diagram corresponding to the first label on
the arrow is shown.
Note that for each diagram, the minimal value of $\hel$ precisely
matches the number of non-MHV intersections.

\subsection{Resolutions}
\label{sec:resolutions}

In each of the 14 twistor diagrams shown
in~\figRef{seed-amplituhedron-diagrams}, the configuration
manifestly exhibits a total of
$2 \nFull + \nEmpty = 8$ on-shell conditions, where
$\nFull$ is the number of filled nodes and $\nEmpty$ is the number
of empty nodes (including, in each diagram, the node at the intersection
of the two loop momenta).

However, on certain sufficiently high codimension boundaries, additional on-shell conditions can be implied
by the others and are therefore ``accidentally''
satisfied.  This phenomenon occurs for the four
twistor diagrams in~\figRef{seed-amplituhedron-diagrams} that have
been drawn with
a filled node at the point $Z_i$ and two empty nodes
in close proximity
(grouped in a faint gray circle in~\figRef{seed-amplituhedron-diagrams}),
representing the four on-shell conditions
\begin{align}
\label{eqn:resolution}
\langle \mathcal{L}^{(1)}\,i{-}1\,i\rangle =
\langle \mathcal{L}^{(1)}\,i\,i{+}1\rangle =
\langle \mathcal{L}^{(2)}\,i\,i{+}1\rangle =
\langle \mathcal{L}^{(1)}\,\mathcal{L}^{(2)}\rangle = 0\,.
\end{align}
The first three conditions are satisfied by
$\mathcal{L}^{(1)} = (Z_i, A)$ and
$\mathcal{L}^{(2)} = (\alpha Z_i + (1 - \alpha) Z_{i+1}, B)$
for any points $A$, $B$.
Then, for generic $A$ and $B$, the fourth condition
in~\eqnRef{resolution} implies that $\alpha = 1$, so the line
$\mathcal{L}^{(2)}$ is forced to pass through the point $Z_i$.
Therefore, configurations of this type satisfy the additional
on-shell condition $\langle \mathcal{L}^{(2)}\,i{-}1\,i\rangle = 0$.

This phenomenon reflects the fact that in general, the
on-shell conditions satisfied by a given configuration are
not independent: some of them may be implied by the others.
In~\cite{Dennen:2016mdk} it was found that solving the Landau equations
for boundaries of this type was rather subtle, and required first
identifying a suitable minimal subset of independent on-shell conditions,
a process
called \emph{resolution}.
It was suggested that a resolution must satisfy two
criteria:  (1) the chosen subset of on-shell conditions must imply
the full set of conditions satisfied for generic (projected)
external data, and (2) the Landau diagram corresponding to the subset
must be planar.

The example considered above describes a configuration
that satisfies five on-shell conditions, the four shown
in~\eqnRef{resolution} and also $\langle \mathcal{L}^{(2)}\,i{-}1\,i\rangle
= 0$. There are four possible resolutions that satisfy criterion
(1): we can simply omit any one of the conditions
except for $\langle \mathcal{L}^{(1)}\,\mathcal{L}^{(2)}\rangle = 0$.
However not all four choices will satisfy criterion (2), depending
on the points $A$ and $B$.  For the four configurations
appearing in~\figRef{seed-amplituhedron-diagrams} that require
resolution, there are in each case precisely two valid resolutions:
we can omit either $\langle \mathcal{L}^{(2)}\,i{-}1\,i\rangle = 0$
(as was done in~\eqnRef{resolution}), or we can omit
$\langle \mathcal{L}^{(1)}\,i\,i{+}1\rangle = 0$.

In~\figRef{seed-amplituhedron-diagrams} we have chosen to always
draw a resolved configuration in the four cases where it is necessary.
However, in order to avoid clutter we do not draw
both resolutions unless they give rise to inequivalent diagrams.
There are at least three reasons for preferring the resolved configurations.
First of all, it becomes somewhat less
clear how to see the
action of the three graph operators $\ko{}$, $\uo{}$ and
$\ro{}$ on an unresolved configuration.
Also, the need for resolution is an accident that occurs
only when both loop momenta lie in the low-$\hel$ branch of solutions
to their respective on-shell conditions (or, by parity symmetry,
when they both lie in the high-$\hel$ branch).  If one of them
lies in the low-$\hel$ branch and the other lies in the high-$\hel$
branch, then for generic (projected) external data only the
resolved configuration(s) exist; the ``extra'' on-shell condition
would place restrictions on the external data.
Finally, when we turn our attention to finding Landau
singularities in~\secRef{nmhv-landau-analysis},
we will always want to work
with resolved diagrams since these give us the independent sets
of on-shell conditions for which we will need to solve the
Landau equations~\cite{Dennen:2016mdk}.

\subsection{Relaxations}

All lower-codimension $\mathcal{L}$-boundaries
are relaxations:  they can be generated
by releasing one or more of the seven
on-shell conditions (excepting $\langle \mathcal{L}^{(1)}\,
\mathcal{L}^{(2)}\rangle = 0$, which we always preserve)
satisfied on the maximal boundaries.
Boundaries of this type can be generated by acting on the
twistor diagrams
in~\figRef{seed-amplituhedron-diagrams}
with sequences of the graph operators $\uo{}$ and $\ro{}$.
In this way one could imagine uplifting the figure
to a three-dimensional generalization
of Fig.~2 of~\cite{Prlina:2017azl}, with the top layer
being a copy of~\figRef{seed-amplituhedron-diagrams}
showing the maximal codimension boundaries ($d=8$), the next layer
showing those with $d=7$, etc.
One novelty compared to the one-loop analysis of~\cite{Prlina:2017azl}
is that starting at two loops the relaxation of a boundary
is not necessarily still a boundary --- this will only be the case
if the Landau diagram of the relaxation continues to be planar.

Rather than attempting to draw the aforementioned web of
interconnected boundaries in a single figure,
we summarize our results in terms of the corresponding Landau diagrams in
Tabs.~\ref{tab:mhv-results}--\ref{tab:n4mhv-results} grouped
according to the minimum helicity for which the configuration is
valid, i.e.~the minimum $\hel$ for which $\mathcal{A}_{n, \hel, 2}$ has
boundaries of the type shown in the corresponding twistor diagram.
Because the maximal codimension singularities have
$d_1 + d_2 = 7$, the corresponding Landau diagrams always have
the topology of a planar pentagon-box.

%
%
%
\begin{table}
\centering
\begin{tabular}
{
>{\centering\arraybackslash} m{0.035\textwidth}
>{\centering\arraybackslash} m{0.275\textwidth}
>{\centering\arraybackslash} m{0.6\textwidth}
}
& Twistor Diagram & Landau Diagram \\
\hline \hline
(a)
&
\includegraphics{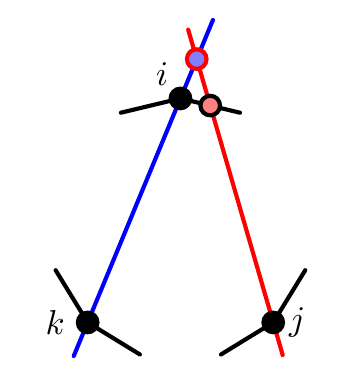}
&
\includegraphics{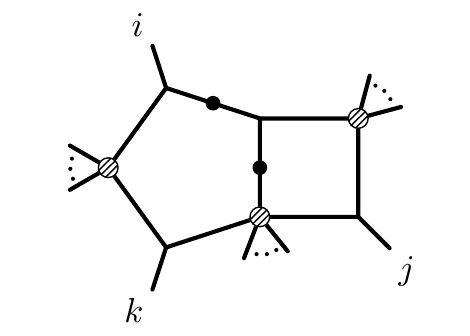}
\\[-9pt]
\end{tabular}
\caption[Results for $\nkmhv{k \ge 0}$]{%
The twistor and Landau diagram describing a type
(the unique type, for $\hel = 0$)
of resolved maximal
codimension boundary of $\nkmhv{\hel \ge 0}$ amplituhedra.
}
 \label{tab:mhv-results}
\end{table}
%
%
%

As mentioned above
the lower codimension singularities can be obtained by acting
on the twistor diagrams with sequences of
$\uo{}$ and $\ro{}$ operators.  As discussed in Sec~5.2
of~\cite{Prlina:2017azl}, at one loop these operators generate relaxations
that respectively preserve or increase, but can never
decrease, the minimum helicity
for which a configuration is valid.
There is however a subtlety with the $\uo{}$ operator at two loops.
Recall that $\uo{i,\mp}$ is the ``unpinning'' operator which acts on a
loop momentum $\mathcal{L}$
passing through some point $Z_i$ by relaxing
the on-shell condition
$\langle \mathcal{L}\, i\, i{\pm}1\rangle = 0$.
This can have the effect of turning what was a manifest intersection
between the two loop momenta into a non-manifest intersection,
which requires increasing the minimum helicity by 1.

%
%
%
\begin{table}
\centering
\begin{tabular}
{
>{\centering\arraybackslash} m{0.035\textwidth}
>{\centering\arraybackslash} m{0.275\textwidth}
>{\centering\arraybackslash} m{0.6\textwidth}
}
& Twistor Diagram & Landau Diagram \\
\hline \hline
(a)
&
\includegraphics{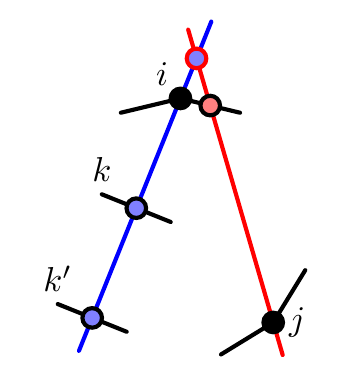}
&
\includegraphics{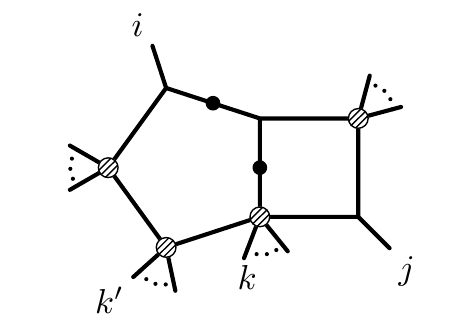}
\\[-9pt]
(b)
&
\includegraphics{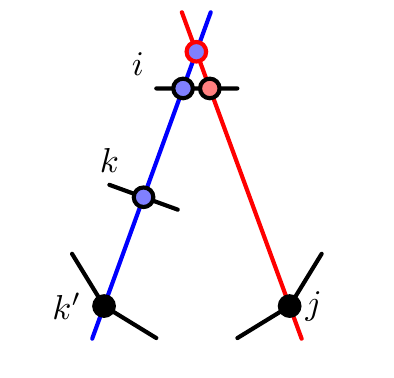}
&
\includegraphics{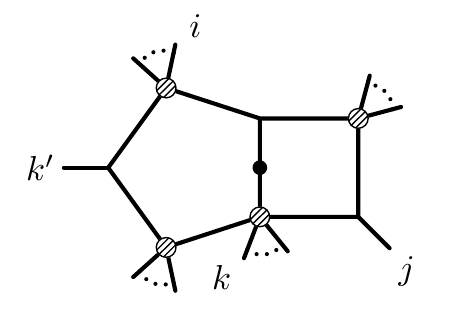}
\\[-9pt]
(c)
&
\includegraphics{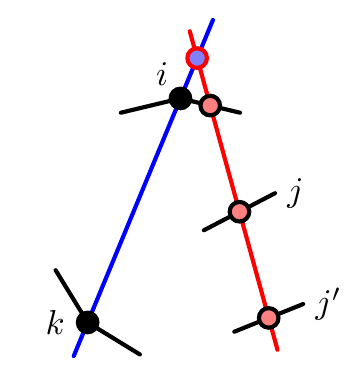}
&
\includegraphics{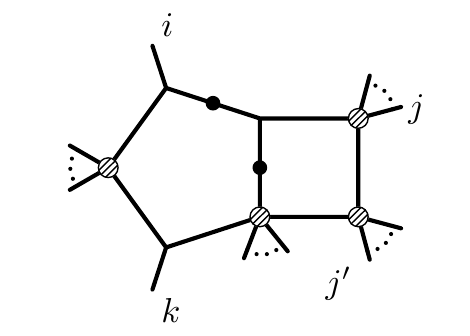}
\\[-9pt]
(d)
&
\includegraphics{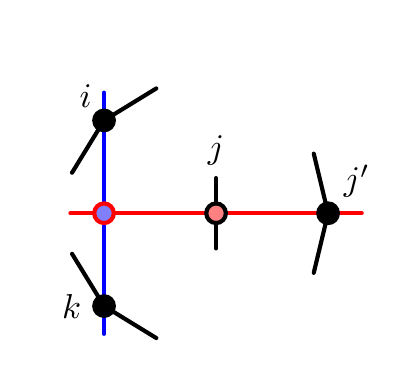}
&
\includegraphics{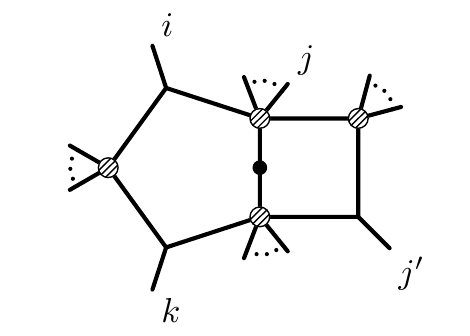}
\\[-9pt]
\end{tabular}
\caption[Results for $\nkmhv{k \ge 1}$]{%
The twistor and Landau diagrams describing types of (resolved,
in (a) and (c)) maximal
codimension boundaries of $\nkmhv{\hel \ge 1}$ amplituhedra.
}
 \label{tab:nmhv-results}
\end{table}
%
%
%

%
%
%
\begin{table}
\centering
\begin{tabular}
{
>{\centering\arraybackslash} m{0.035\textwidth} 
>{\centering\arraybackslash} m{0.275\textwidth} 
>{\centering\arraybackslash} m{0.6\textwidth}  
}
& Twistor Diagram & Landau Diagram \\
\hline \hline
(a)
&
\includegraphics{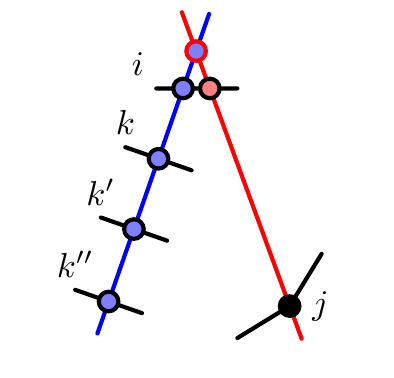}
&
\includegraphics{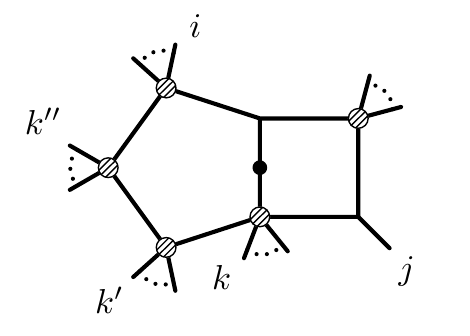}
\\[-9pt]
(b)
&
\includegraphics{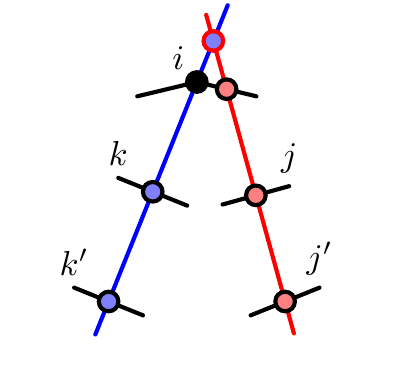}
&
\includegraphics{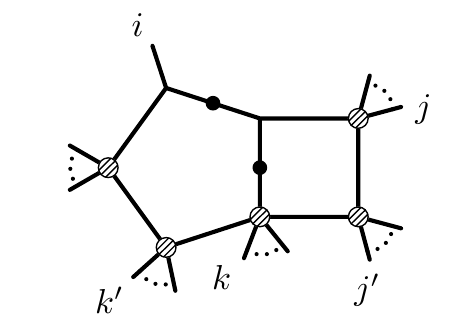}
\\[-9pt]
(c)
&
\includegraphics{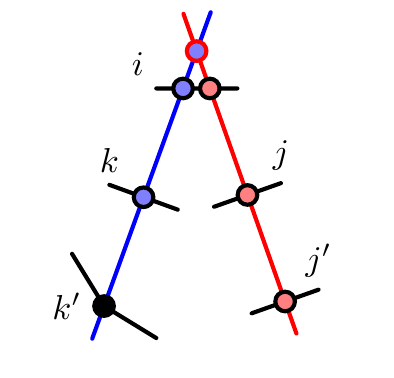}
&
\includegraphics{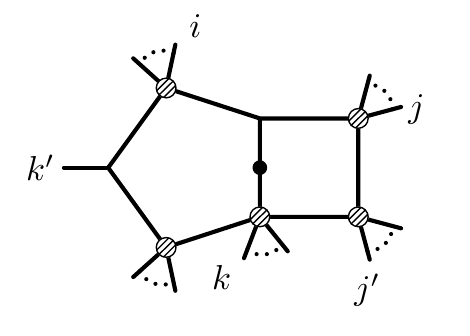}
\\[-9pt]
(d)
&
\includegraphics{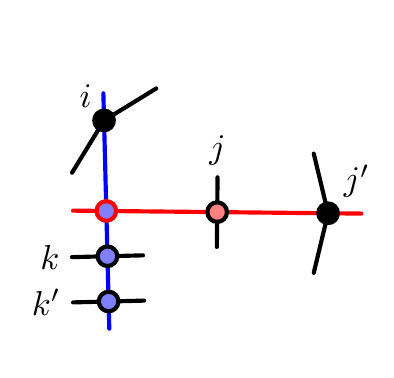}
&
\includegraphics{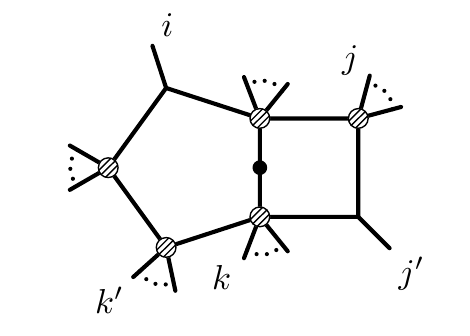}
\\[-9pt]
(e)
&
\includegraphics{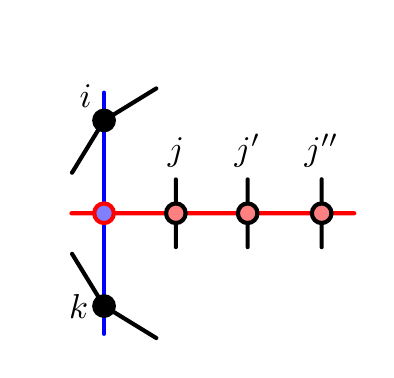}
&
\includegraphics{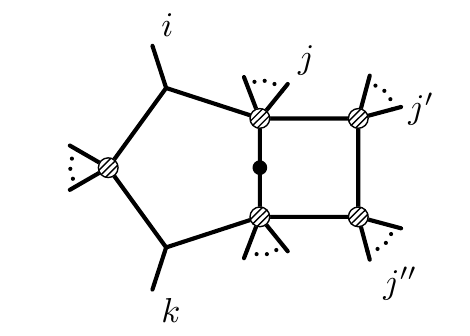}
\\[-9pt]
\end{tabular}
\caption[Results for $\nkmhv{k \ge 2}$]{%
The twistor and Landau diagrams describing types of (resolved,
in (b)) maximal
codimension boundaries of $\nkmhv{\hel \ge 2}$ amplituhedra.
}
 \label{tab:n2mhv-results}
\end{table}
%
%
%

In the tables we have introduced a new graphical notation in order
to account for this phenomenon:  a propagator with a black dot
denotes an on-shell condition that cannot be relaxed without
increasing the minimum helicity for which the configuration is valid.
(We also always draw a black dot on the $\langle
\mathcal{L}^{(1)}\,\mathcal{L}^{(2)}\rangle$ propagator,
as a reminder that we never want to relax it.)
Consider for example the twistor diagram
in~\tabRef{mhv-results}(a). The two loop momenta manifestly
intersect at the point $Z_i$ as explained in the previous section,
but this will no longer be the case if we act on this twistor
diagram with $\uo{i,-}$.  Instead, the configuration would
become NMHV rather than MHV (in fact, it would become a
relaxation of~\tabRef{nmhv-results}(d), up to relabeling).
For this reason we draw a black dot on the $(i\,i{+}1)$
propagator on the pentagon in the Landau diagram
of~\tabRef{mhv-results}(a).

%
%
%
\begin{table}
\centering
\begin{tabular}
{
>{\centering\arraybackslash} m{0.05\textwidth} 
>{\centering\arraybackslash} m{0.3\textwidth} 
>{\centering\arraybackslash} m{0.35\textwidth}  
}
& Twistor Diagram & Landau Diagram\\
\hline \hline
(a)
&
\includegraphics{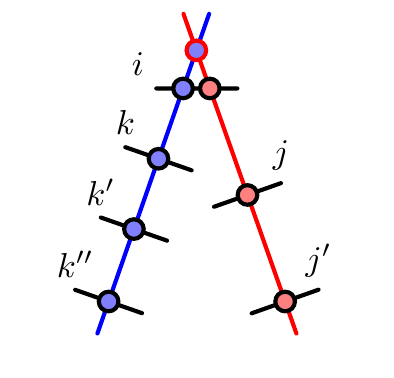}
&
\includegraphics{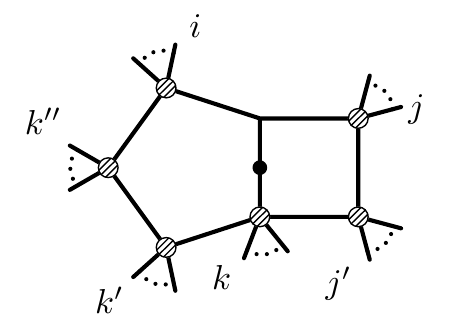}
\\[-9pt]
(b)
&
\includegraphics{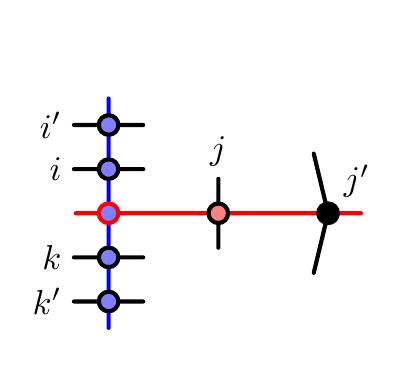}
&
\includegraphics{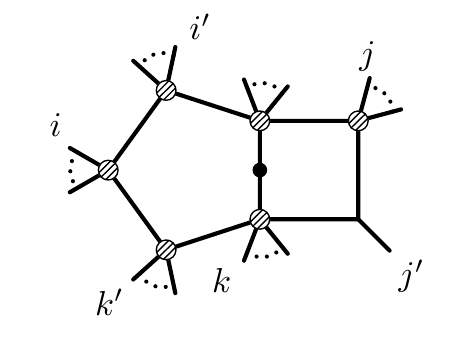}
\\[-9pt]
(c)
&
\includegraphics{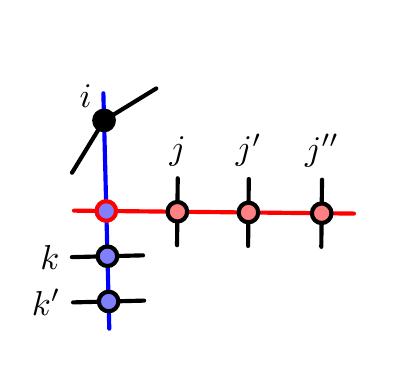}
&
\includegraphics{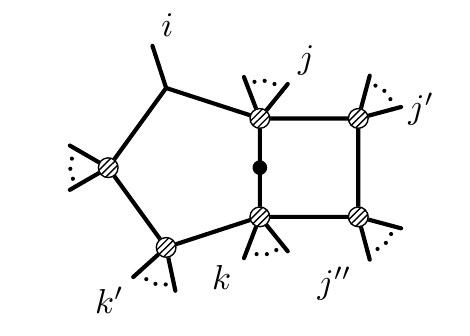}
\\[-9pt]
\end{tabular}
\caption[Results for $\nkmhv{k \ge 3}$]{%
The twistor and Landau diagrams describing types of
maximal
codimension boundaries of $\nkmhv{\hel \ge 3}$ amplituhedra.
}
 \label{tab:n3mhv-results}
\end{table}
%
%
%

\subsection{Closing Comments}
\label{sec:closing}

In summary,
to get the full list of Landau diagrams at
helicity $\hel = 0, 1, 2, 3, 4$, one must therefore
consider all of the Landau diagrams in
Tables.~\ref{tab:mhv-results} through~\ref{tab:n4mhv-results},
respectively, together with the diagrams generated therefrom by collapsing
any subset of undotted propagators.

In~\figRef{seed-amplituhedron-diagrams} and in the tables
we have chosen to always
draw the loop momentum satisfying $d_1 = 4$
in blue and the one satisfying $d_2 = 3$ in red, but of course
the amplituhedron is symmetric under the exchange of any $\mathcal{L}$'s
so
in each case
both assignments \color{blue} $\mathcal{L}^{(1)}$\color{black},
\color{red} $\mathcal{L}^{(2)}$ \color{black} and
\color{blue} $\mathcal{L}^{(2)}$\color{black},
\color{red} $\mathcal{L}^{(1)}$\color{black}
describe valid boundaries.

The Landau diagrams in Tables~\ref{tab:mhv-results}--\ref{tab:n4mhv-results}
are always drawn with the understanding that all indicated labels
are cyclically ordered: $i < i' < j < j' < j'' < k < k' < k'' < i$ (mod $n$).
However, the ordering of intersections along the red or blue
loop momentum lines carries no significance.  Therefore,
as described in Sec.~5.1 of~\cite{Prlina:2017azl}, there is a second
type of ambiguity between the two classes of diagrams.
For example, the twistor diagram in Tab.~\ref{tab:nmhv-results}(a)
is agnostic about the cyclic ordering of $i$, $k$, and $k'$;
the two independent choices lead to the Landau diagram shown in the table
or to its mirror image.
In all of the tables we use primes (and, when necessary, also
double primes) to indicate pairs (or triplets)
of nodes that can be exchanged, as far as the
twistor diagram is concerned.
Sometimes, as
in the example
Tab.~\ref{tab:nmhv-results}(a) just considered, an exchange generates
a Landau diagram of the same topology, but in other cases it can generate
a new topology.  For example, exchanging $k$ and $k'$
in the twistor diagram of Tab.~\ref{tab:nmhv-results}(b) generates
the new Landau diagram
\begin{align}
\begin{gathered}
\includegraphics{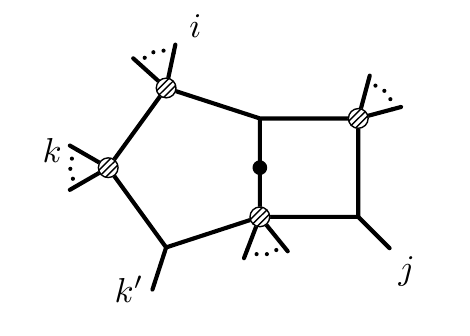}
\end{gathered}\,,
\nonumber
\end{align}
where it is to be understood that $i < j < k' < k < i$.

%
%
%
\begin{table}
\centering
\begin{tabular}
{
>{\centering\arraybackslash} m{0.05\textwidth} 
>{\centering\arraybackslash} m{0.3\textwidth} 
>{\centering\arraybackslash} m{0.35\textwidth}  
}
 & Twistor Diagram & Landau Diagram \\
\hline \hline
(a)
&
\includegraphics{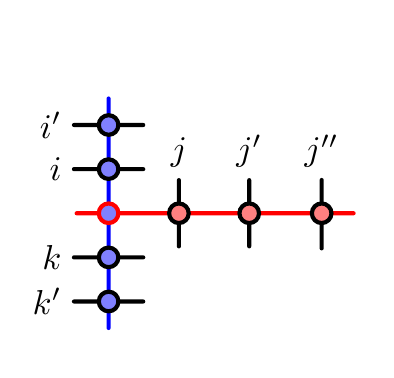}
&
\includegraphics{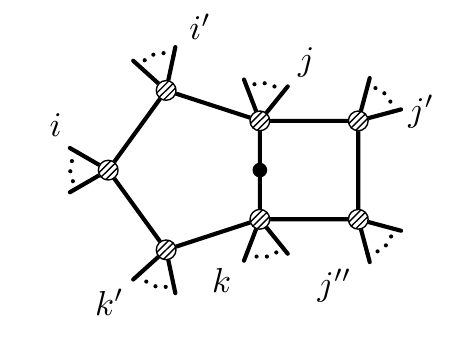}
\\[-9pt]
\end{tabular}
\caption[Results for $\nkmhv{k \ge 4}$]{%
The twistor and Landau diagram describing a type of
maximal codimension boundary of $\nkmhv{\hel \ge 4}$ amplituhedra.
}
 \label{tab:n4mhv-results}
\end{table}
%
%
%

Let us also note
that although when interpreted literally as configurations of
intersecting lines in $\mathbb{P}^3$ most twistor diagrams only
depict the low-$\hel$ branch of solutions to a given set of
on-shell conditions, it is clear that additional, higher-$\hel$
boundaries can be generated by replacing one or both of the
$\mathcal{L}$'s with their parity conjugates.  The twistor
diagrams appearing in~\figRef{seed-amplituhedron-diagrams}
and in the five tables can therefore
each be thought of as representing four different types of boundaries
corresponding to the same Landau diagram.

Finally, we detail, in Appendix~\ref{sec:twistor-to-landau}, how a partial edge-to-node duality
maps between the twistor diagrams on the left
and the Landau diagrams on the right of these tables, when the two diagrams are treated as graphs.
On the one hand, it is not surprising that there exists some map between
these two classes of graphs, since both are designed to encode the same information.
On the other hand, it is intriguing that there is a straightforward map
between a generic Landau diagram and the minimum-helicity solution
to the on-shell conditions of said diagram
in the very particular choice of loop momentum twistor coordinates.
This observation is also reminiscent of the map from Feynman integrals to their duals
that aided in exploring the dual conformal invariance of SYM theory
amplitudes~\cite{Drummond:2006rz,Alday:2007hr,Drummond:2008vq}
but here, enticingly, this partial edge-to-node map is well-defined even on nonplanar graphs.

\section{The Connection With On-Shell Diagrams}
\label{sec:on-shell-diagrams}

So far, we have seen that to each boundary of an amplituhedron one
can associate a Landau diagram
which encodes information about the singularities of the associated
amplitude.
In this section we explore the connection between Landau diagrams
and a class of closely related diagrams that also encode information about
an amplitude's mathematical structure:  the on-shell
diagrams of~\cite{ArkaniHamed:2012nw}.
We explain and demonstrate in several examples that
for a given amplitude, the information content of
certain on-shell diagrams matches the combined
information content in the amplituherdon and Landau diagrams.
Except possibly for cases of the type discussed in the paragraph
following \eqnRef{toreject}, we expect our arguments
to also hold for amplitudes at higher loop order
and higher helicity.

One reason to shift focus to on-shell diagrams is that
anything that can be formulated in terms of
the on-shell diagrams discussed here potentially generalizes
to more general quantum field theories including less supersymmetric
theories as well as the full, non-planar super-Yang--Mills theory.
The major difference is that in the planar theory,
the relevant Landau diagrams can, in principle, be read
off from the boundaries of $\mathcal{A}_{n,\hel,L}$ for
arbitrary $n$, $\hel$, and $L$, while  in the non-planar
sector there is currently no known supplier of this list of diagrams.
Nevertheless, assuming one has a way to generate a representation for
a given non-planar amplitude in terms of Feynman integrals,
all of the techniques discussed in this section
apply equally well to those non-planar integrals.

Putting that ambitious motivation aside, in the rest of this section
we stick to planar SYM theory and show in several examples
that a given Landau diagram encodes a singularity of
an $\nkmhv{\hel}$ amplitude only if the diagram can be decorated
in such a way that it becomes an on-shell diagram
associated with an $\nkmhv{\hel}$ amplitude.
We begin with a brief review of on-shell diagrams.

\subsection{On-Shell Diagrams}
\label{sec:on-shell-diagrams-review}

An~\emph{on-shell diagram}, as introduced in~\cite{ArkaniHamed:2012nw},
is a connected trivalent graph
with each node having one of two distinct decorations,
traditionally denoted by
coloring them black or white.
In the application to scattering amplitudes,
each edge of the diagram represents an on-shell condition
(just like in a Landau diagram) and
each black (white) node corresponds to a three-point MHV
($\overline{\rm MHV}$)
tree-level superamplitude.
A straightforward generalization allows nodes of higher degree
which represent higher-point tree-level superamplitudes.
These we depict by a shaded node.

We refer the reader to~\cite{ArkaniHamed:2012nw} for details,
recalling here only a few basic facts.
A \emph{tree-level superamplitude} of
\emph{Grassmann weight} $\kappa$ is a rational function
of (projected) external data that is a homogeneous polynomial
of degree $4 \kappa$ in certain Grassmann variables (the fermionic
partners of the momentum twistors $Z_i$).
Three-point MHV and $\overline{\rm MHV}$ amplitudes respectively
have $\kappa = 2$ and $\kappa= 1$ while for $n > 3$ an $n$-point
amplitude with helicity $\hel$ has $\kappa = \hel+2$.
To each on-shell
diagram there is an associated differential form that is obtained
by first multiplying together
the tree-level superamplitudes represented by each of the diagram's
nodes, and then sewing them together according to a set of simple
rules
that involve integrating over four Grassmann variables for each
internal edge (propagator) in the diagram.
Such forms are the values of the residue of the amplitude's
integrand at specific
loci in loop momentum space.

Consider an on-shell diagram $\delta$.  Let $\iota$ be the number
of internal edges of $\delta$, and for each node $\nu$ let $\kappa_\nu$
be the Grassmann weight of the tree-level superamplitude at $\nu$.
As a result of the rules just reviewed,
the total Grassmann weight of $\delta$ is
\begin{align}
\label{eqn:total-grassmann-weight}
\kappa_\delta = \sum_{\nu} \kappa_\nu - \iota \,,
\end{align}
and the total helicity is $\hel_\delta = \kappa_\delta-2$.

To assign a~\emph{coloring} to a Landau diagram depicting some
set of on-shell conditions means to assign to each trivalent node
in the diagram either a white or black coloring, and to
assign to each node $\nu$ of degree $n > 3$ some helicity
$\hel_\nu = \kappa_\nu - 2 \in \{0, \ldots, n - 4\}$.
Since $\iota$ is fixed by the propagator structure of the diagram,
and each $\kappa_\nu$ is positive,
it is clear from~\eqnRef{total-grassmann-weight}
that the minimal Grassmann weight of a given Landau diagram
results from coloring all trivalent nodes white and from
assigning all nodes of higher degree to be MHV ($\kappa_\nu = 2$).
In this way we see that
the Grassmann weight of an arbitrary coloring of a given Landau diagram
is bounded below
by
\begin{align}
\label{eqn:minimum-grassmann}
\kappa \ge \kappa_{\textrm{min}} = \ntri + 2 \nhigh - \iota \,,
\end{align}
where $\ntri$ is the number of trivalent nodes and
$\nhigh$ is the number of nodes of degree higher than three.
This implies a minimal helicity
sector $k_{\textrm{min}} = \kappa_{\textrm{min}} - 2$ for which the
Landau diagram can be relevant.

If a diagram has $\ntri$ trivalent indices,
there are $2^{\ntri}$ colorings of the trivalent nodes, but
in general some of these may lead to on-shell diagrams that
evaluate to zero.
In practice we count the number of permissible colorings of a diagram
by solving the on-shell
conditions implied by the diagram
and mapping each resulting solution to a specific coloring
(see~\cite{ArkaniHamed:2012nw}).
As discussed in~\cite{ArkaniHamed:2010gh}, solving a
set of on-shell conditions in momentum twistor space
amounts to solving a Schubert problem.
At one loop these problems have in general two solutions,
while for an $L$-loop Landau diagram we would in general
expect $2^L$ branches of solutions.
Given a solution to a Schubert problem in momentum twistor space,
it is straightforward\footnote{We thank J.~Bourjaily
for explaining this point to us.}
to check if a given trivalent node is MHV or $\overline{\rm MHV}$
by considering
the rank of the three momentum-twistor lines at the node.
For an MHV node, the three twistors have full rank,
while for an $\overline{\rm MHV}$ node the rank is less than full.
This process is illustrated explicitly in several examples
in the following section.

In summary, we have reviewed that a given Landau diagram
encodes a set of on-shell conditions, and the various branches
of solutions
to those conditions
correspond in general to different minimum helicity sectors.
The permissible colorings of a Landau diagram
are in one-one correspondence with those branches,
and the Grassmann weight $\kappa$ of each
such
Landau-turned-on-shell diagram is related to the minimum helicity sector
$\hel$
of the corresponding solution via $\kappa=\hel+2$.

This observation provides an alternative way to phrase
the Landau-equation-based
algorithm we employ to identify singularities of amplitudes,
compared for example to the way it is phrased
in the conclusion of~\cite{Dennen:2016mdk}
or in Sec.~2.5 of~\cite{Prlina:2017azl}.
For one thing, it means we can
identify a singularity of a Landau diagram
as a singularity of $\nkmhv{\hel}$ amplitudes only if the
diagram admits a coloring with total helicity $\hel$ (equivalently,
Grassmann weight $\hel+2$).
More specifically, when first solving the on-shell conditions
(a subset of the Landau equations) for a given
Landau diagram, each solution directly indicates,
via the test reviewed in the previous paragraph, the helicity
sector for which the singularity associated to that solution is relevant.
In the on-shell diagram approach this step is the analog in
the amplituhedron approach of
identifying the values of $\hel$ for which the momentum twistor
solution lies on the boundary of the $\nkmhv{\hel}$ amplituhedron.
In the amplituhedron-based approach, there is potential for
confusion because solving the Kirchhoff conditions (the remaining
Landau equations) can lead to solutions for loop momenta
that lie outside the $\nkmhv{\hel}$ amplituhedron.  The on-shell
diagram approach bypasses this confusion because the Kirchhoff
conditions only further localize a loop momentum solution whose helicity
sector has already been identified.

\subsection{Examples at One and Two Loops}
\label{sec:on-shell-diagram-examples}

We now consider several examples in order to emphasize the following point:
\begin{align}
\begin{array}{l}
\textrm{A
Landau diagram contributes singularities to an $\nkmhv{\hel}$ amplitude
}
\\
\textrm{only
if the diagram permits a coloring with total Grassmann weight $\hel+2$.
}
\end{array}
\end{align}
For each of our examples, we also list the values of the loop momenta
corresponding to the colorings of the correct Grassmann weight. For the
one-loop examples the same information can be read off from Tab.~1
of~\cite{Prlina:2017azl}. We will show how the on-shell diagram and
amplituhedron-based methods work in tandem to quickly identify the
helicity sector
for which a given solution to the set of on-shell conditions is relevant.

\subsubsection*{One-loop Two-mass Easy Box}

The on-shell conditions
\begin{align}
\langle \mathcal{L}\,i{-}1\,i\rangle =
\langle \mathcal{L}\,i\,i{+}1\rangle =
\langle \mathcal{L}\,j{-}1\,j\rangle =
\langle \mathcal{L}\,j\,j{+}1\rangle = 0
\end{align}
admit two solutions, called branches (12) and
(13) in~\cite{Prlina:2017azl}.
In~\tabRef{two-mass-easy-colorings} we pair the momentum
twistor representation of each solution
with the associated on-shell diagram, i.e.~colored Landau diagram.
Having this information accessible will prove useful when considering
two loops.

In~\tabRef{two-mass-easy-colorings}(a) the minimum Grassmann weight is
computed according to~\eqnRef{minimum-grassmann} and found to be
\begin{align}
\kappa_{\textrm{min}} =
\underbrace{1+1}_{\textrm{white}} + \underbrace{2 + 2}_{\textrm{higher}}-4
= 2
\end{align}
so that it is an MHV ($\hel=2-2=0$) coloring.

In~\tabRef{two-mass-easy-colorings}(b) the minimum Grassmann weight is
\begin{align}
\kappa_{\textrm{min}} =
\underbrace{2+2}_{\textrm{black}} + \underbrace{2 + 2}_{\textrm{higher}}-4
= 4
\end{align}
so that it is an $\nkmhv{2}$ ($\hel=4-2=2$) coloring.

Let us now show how to compute
the appropriate node colorings directly from
the momentum twistor solutions in~\tabRef{two-mass-easy-colorings}.
Consider the trivalent node where external label $i$ connects to the loop.
The three lines in momentum twistor space defining the trivalent node are
$(i{-}1\,i)$, $(i\,i{+}1)$, and $\mathcal{L}_*$,
where $\mathcal{L}_*$ is either $(i\,j)$
or $\overline{i} \cap \overline{j}$.
Taking first $\mathcal{L}_{*} = (i\,j)$ ,
we seek the dimension of the space spanned by the three momentum-twistor
lines.
One way to compute this is to ask for the rank of the matrix:
\begin{align}
\textrm{rank}
\left(
\begin{array}{c}
(i{-}1\,i) \\
(i\,i{+}1) \\
(i\,j)
\end{array}
\right)
=
\textrm{rank}
\bordermatrix{
& i{-}1 & i & i{+}1 & j \cr
& 1 & 0 & 0 & 0 \cr
& 0 & 1 & 0 & 0 \cr
& 0 & 1 & 0 & 0 \cr
& 0 & 0 & 1 & 0 \cr
& 0 & 1 & 0 & 0 \cr
& 0 & 0 & 0 & 1 \cr
 }
 =
 4
\end{align}
which has maximal rank. So the node is MHV, and colored white.

In contrast, consider the other solution $\mathcal{L}_{*} =
\overline{i} \cap \overline{j}$.
The analogous matrix is then
\begin{align}
\textrm{rank}
\left(
\begin{array}{c}
(i{-}1\,i) \\
(i\,i{+}1) \\
\overline{i} \cap \overline{j}
\end{array}
\right)
=
\textrm{rank}
\bordermatrix{
& i{-}1 & i & i{+}1 \cr
& 1 & 0 & 0 \cr
& 0 & 1 & 0 \cr
& 0 & 1 & 0 \cr
& 0 & 0 & 1 \cr
& \langle i\,\overline{j} \rangle & - \langle i{-}1\, \overline{j}
\rangle & 0 \cr
& 0 & \langle i{+}1 \overline{j} \rangle &
- \langle i \, \overline{j} \rangle \cr
 }
 =
 3
\end{align}
which does not have maximal rank.
Thus the second solution is encoded in an $\overline{\rm MHV}$ node
at $i$, colored black.
The colorings of the node at $j$ can be computed analogously.

%
%
%
\begin{table}
\centering
\begin{tabular}
{
>{\centering\arraybackslash} m{0.05\textwidth}
>{\centering\arraybackslash} m{0.2\textwidth}
>{$}c<{$} 
>{$}c<{$} 
}
& Coloring & \nkmhv{\hel} & \textrm{Twistor Solution}
\\
\hline \hline
(a)
&
\includegraphics{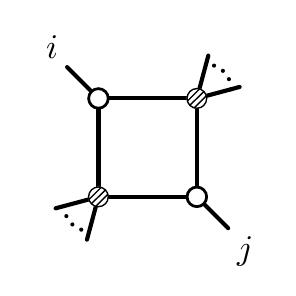}
&
\hel \ge 0
&
\mathcal{L} = (i\,j)
\\
(b)
&
\includegraphics{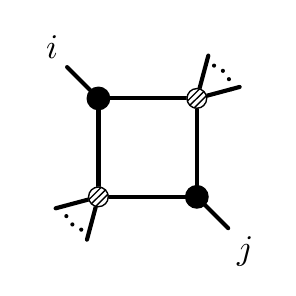}
&
\hel \ge 2
&
\mathcal{L} = \overline{i} \cap \overline{j}
\\
\end{tabular}
\caption[Two colorings of the two-mass easy box.]{%
Two colorings of the two-mass easy box.
Row (a) shows the MHV coloring and momentum twistor solution to the
on-shell conditions,
and row (b) shows the same for the $\nkmhv{2}$ solution.
}
 \label{tab:two-mass-easy-colorings}
\end{table}
%
%
%

%
%
%
\begin{table}
\centering
\begin{tabular}
{
>{\centering\arraybackslash} m{0.05\textwidth}
>{\centering\arraybackslash} m{0.2\textwidth}
>{$}c<{$} 
>{$}c<{$} 
}
& Coloring & \nkmhv{\hel} & \textrm{Twistor Solution}
\\
\hline \hline
(a)
&
\includegraphics{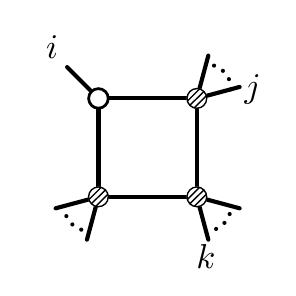}
&
\hel \ge 1
&
\mathcal{L} = (i\,j\,j{+}1) \cap (i\,k\,k{+}1)
\\
(b)
&
\includegraphics{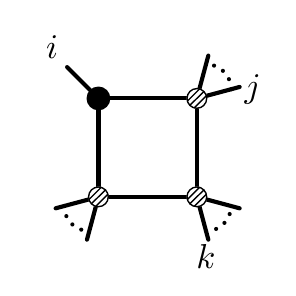}
&
\hel \ge 2
&
\begin{array}{c}
\mathcal{L}= (A\,B)  \\
A = (j\,j{+}1) \cap \overline{i}  \\
B = (k\,k{+}1) \cap \overline{i}
\end{array}
\\
\end{tabular}
\caption[Two colorings of the two-mass easy box.]{%
Two colorings of the three-mass box.
Row (a) shows the NMHV coloring and
momentum twistor solution to the on-shell conditions,
and row (b) shows the same for the $\nkmhv{2}$ solution.
}
 \label{tab:three-mass-box-colorings}
\end{table}
%
%
%

\subsubsection*{One-loop Three-mass Box}

We perform the same exercise for the three-mass box on-shell
conditions
\begin{align}
\langle \mathcal{L}\, i{-}1\,i\rangle =
\langle \mathcal{L}\,i\,i{+}1\rangle =
\langle \mathcal{L}\,j\,j{+}1\rangle =
\langle \mathcal{L}\,k\,k{+}1\rangle = 0\,.
\end{align}
The
solutions
of the on-shell conditions
are matched to the two on-shell diagram colorings
in~\tabRef{three-mass-box-colorings},
and the corresponding
minimum Grassmann weights are computed using~\eqnRef{minimum-grassmann}.
The colorings are also directly calculable from the momentum twistor
solutions as in the previous two-mass easy box example.
The three-mass box is worth pointing out because in this
case neither coloring is MHV, in contrast to the previous
example.

\subsubsection*{Two-loop Pentagon-box}

We can recycle our knowledge of one-loop solutions to determine the helicity
sectors to which a given two-loop Landau diagram contributes its singularities.
We consider the pentagon-box of~\tabRef{nmhv-results}(a) as an exemplar.
We solve the pentagon-box on-shell conditions as follows.
We first solve the subsystem of four propagators that depend
on only $\mathcal{L}^{(2)}$:
\begin{align}
\label{eqn:three-mass-color-cut}
 \langle \mathcal{L}^{(2)} \, i{-}1 \, i \rangle =
\langle \mathcal{L}^{(2)} \, i \, i{+}1 \rangle =
\langle \mathcal{L}^{(2)} \, k \, k{+}1 \rangle =
\langle \mathcal{L}^{(2)} \, k' \, k'{+}1 \rangle = 0
\end{align}
using either of the two three-mass box solutions
shown in~\tabRef{three-mass-box-colorings}, after an appropriate
exchange of the external labels in order to
match to~\eqnRef{three-mass-color-cut}.
This means there are two branches of colorings: one were the
trivalent node at $i$ is white,
and one where it is black. The two corresponding
solutions $\mathcal{L}^{(2)}_*$ are shown in the first row
of~\tabRef{pentagon-box-colorings}.
For each choice of $\mathcal{L}^{(2)}_*$ we then solve the
remaining four on-shell conditions
\begin{align}
\label{eqn:two-mass-easy-color-cut}
 \langle \mathcal{L}^{(1)} \, i \, i{+}1 \rangle =
\langle \mathcal{L}^{(1)} \, j{-}1 \, j \rangle =
\langle \mathcal{L}^{(1)} \, j \, j{+}1 \rangle =
\langle \mathcal{L}^{(1)}\, \mathcal{L}^{(2)}_* \rangle\,.
\end{align}
These four conditions
constitute a two-mass easy box problem, so we can
utilize~\tabRef{two-mass-easy-colorings}
to identify
the two solutions $\mathcal{L}_*^{(1)}$, which color the trivalent
nodes of the box either both white or both black.
These two solutions
are tabulated in the first column of~\tabRef{pentagon-box-colorings}.
Altogether the table shows a grid containing a total of four
distinct solutions, and the four associated distinct colorings.
From this analysis we conclude that only the solution
\begin{align}
\mathcal{L}^{(2)}_{*,1} =
(i\,k\,k{+}1) \cap (i\,k'\,k'{+}1)\,, \ \mathcal{L}^{(1)}_{*,1}=
(j\,i\,i{+}1) \cap (j\,\mathcal{L}_{*,1}^{(2)}) = (i \, j)
\label{eqn:nmhvsolution}
\end{align}
shown in the top left of~\tabRef{pentagon-box-colorings}
is relevant to the NMHV sector.
This means that
when we turn in the following section to the problem of finding
singularities of NMHV amplitudes by solving the Landau equations,
we can disregard the other three solutions.
Were we to attempt an amplituhedron-based answer to this same question,
we would find that the other
solutions to the on-shell conditions do not lie on a boundary
of $\mathcal{A}_{n,1,2}$.

%
%
%
\begin{table}
\centering
\begin{tabular}
{
>{$}c<{$} 
|>{\centering\arraybackslash} m{0.33\textwidth}
|>{\centering\arraybackslash} m{0.33\textwidth}
}
\mathcal{L}^{(1)}_* \backslash \mathcal{L}^{(2)}_*
 &  $(i\,k\,k{+}1) \cap (i\,k'\,k'{+}1) $ &
 $
 \begin{array}{c}
((k\,k{+}1) \cap \overline{i}\, (k'\,k'{+}1) \cap \overline{i}) \\
\end{array}
 $
\\ \hline
(i\,j)
&
\includegraphics{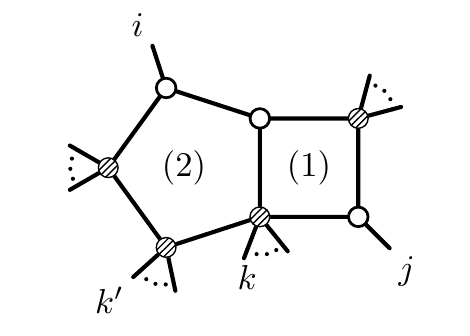}
&
\includegraphics{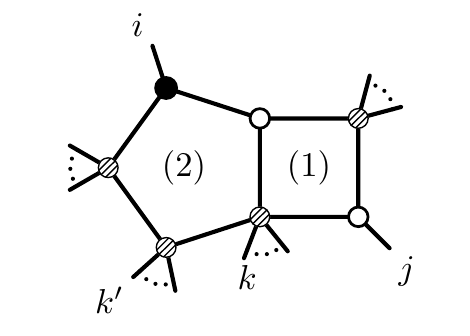}
\\[-12pt]
& $\hel \ge 1$ & $\hel \ge 2 $
\\ \hline
\bar{i} \cap \bar{j}
&
\includegraphics{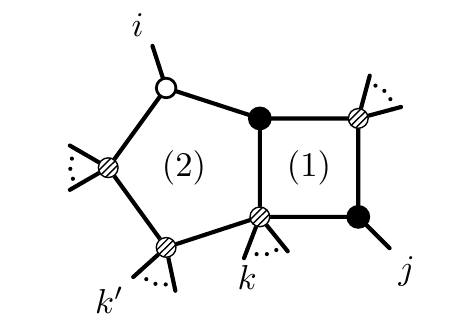}
&
\includegraphics{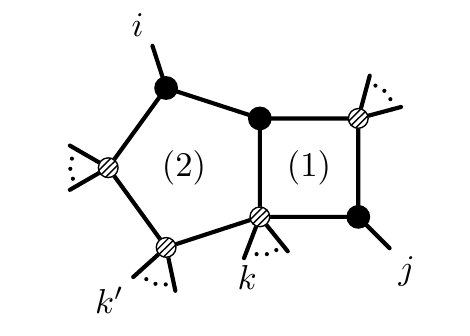}
\\[-12pt]
& $\hel \ge 3$ & $ \hel \ge 4 $ \\
\end{tabular}
\caption[Four colorings of the two-loop pentagon-box.]{%
All permissible colorings of the
trivalent nodes of the
two-loop pentagon-box Landau diagram from~\tabRef{nmhv-results}(a).
The first row shows the two possible solutions
to the three-mass
on-shell conditions
(\eqnRef{three-mass-color-cut}) satisfied by
$\mathcal{L}^{(2)}$, the loop momentum in the pentagon.
The first column shows the two possible solutions
to the two-mass easy on-shell
conditions (\eqnRef{two-mass-easy-color-cut}) satisfied
by $\mathcal{L}^{(1)}$, the loop momentum in the box.
The cell at the intersection of a row and a column is the
colored Landau diagram
that results from the two solutions.
Also indicated in each cell is the minimum helicity sector
of the colored Landau diagram, which is achieved only if the gray
nodes are taken to be MHV.
}
 \label{tab:pentagon-box-colorings}
\end{table}
%
%
%

\subsection*{General Two-Loop Pentagon-Boxes}

By using the same simple counting arguments applied
to the results in Tables~\ref{tab:mhv-results}--\ref{tab:n4mhv-results}, it
is a straightforward
exercise to show that
\begin{itemize}
\item the set of Landau diagrams corresponding to the maximal codimension boundaries of $\mathcal{A}_{n,\hel,2}$ and
\item the set of on-shell diagrams of pentagon-box
topology that admit an $\nkmhv{\hel}$ coloring
\end{itemize}
are the same.
Specifically, the second set may be constructed by starting
with a pentagon-box diagram with no external edges or coloring,
then placing all possible combinations of massive and massless edges on
nodes of the diagram in all possible ways,
and finally enumerating all colorings of the resulting Landau diagrams to
identify the minimum possible value of $\hel$.

\section{Landau Singularities of Two-Loop NMHV Amplitudes}
\label{sec:nmhv-landau-analysis}

Finally we come to step 2 of the algorithm summarized
in Sec.~2.5 of~\cite{Prlina:2017azl}:
in order to determine the locations of Landau singularities of
the two-loop $\nkmhv{\hel}$ amplitude in SYM theory, we must identify,
for each $\mathcal{L}$-boundary of $\mathcal{A}_{n,\hel,2}$
tabulated in~\secRef{presentation},
the codimension-one loci (if there are any) in $\Conf_n(\mathbb{P}^3)$
on which the corresponding Landau equations admit nontrivial solutions.

The ultimate aim of this project has been to derive (or at least to
conjecture)
symbol alphabets for two-loop amplitudes.
However, as discussed in Sec.~7 of~\cite{Prlina:2017azl},
guessing a symbol alphabet from a list of singularity loci
can require a nontrivial extrapolation.
At one loop the extrapolation is straightforward for all Landau diagrams
except the four-mass box.

%
%
%
\begin{figure}
\centering
\begin{tabular}
{
>{\centering\arraybackslash} m{0.3\textwidth} 
>{\centering\arraybackslash} m{0.03\textwidth} 
>{\centering\arraybackslash} m{0.3\textwidth} 
}
\includegraphics{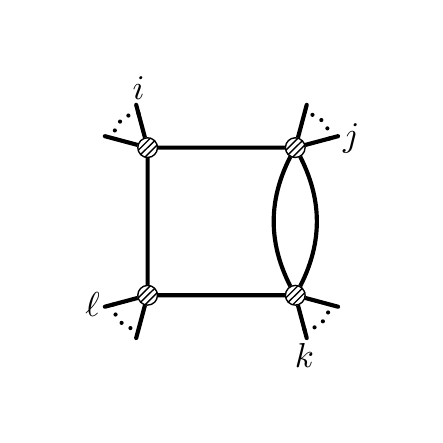}
&
$\sim$
&
\includegraphics{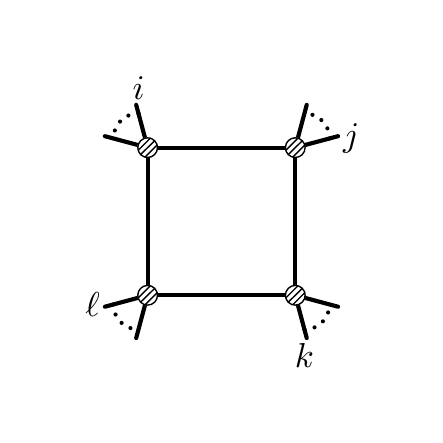}
\end{tabular}
\caption[Four-mass bubble-box is a four-mass box.]{%
The Landau equations of a Landau diagram containing a bubble are
identical to the equations of a Landau diagram with one propagator
of the bubble removed. The two-loop four-mass bubble-box on the left is the only
Landau diagram with a four-mass box contributing to the branch points of the
NMHV amplitude. It has the same well-known branch points as the one-loop four-mass box on the right.
}
\label{fig:bubble-box}
\end{figure}
%
%
%

At two loops, four-mass box subdiagrams
become prevalent starting at $\hel = 2$, where they appear in the maximal
codimension Landau diagram shown
in~\tabRef{nmhv-results}(e), as well as in many of the relaxations of
the other Landau diagrams in~\tabRef{nmhv-results}.
At $\hel=1$
there is a single four-mass bubble-box Landau diagram, \figRef{bubble-box},
relevant to two-loop NMHV amplitudes.
As shown in the Appendix of~\cite{Dennen:2016mdk}, Landau diagrams
containing bubble subdiagrams are
equivalent to the same diagram with one of the propagators of the bubble removed.
So we expect the one-loop four-mass box singularity to reappear
as a singularity of the two-loop NMHV amplitude.
Though we are only guaranteed from this analysis that the singularities
match, we can throw caution to the wind and conjecture that
the same symbol entries that appear in the one-loop four-mass box integral
appear also in two-loop NMHV amplitudes.
Of note here: the four-mass box has support starting at $\hel=2$,
so there is shared singularity structure between the two-loop $\nkmhv{}$ and one-loop $\nkmhv{2}$
amplitudes.

Having dealt with this single caveat,
we restrict our analysis to the remaining NMHV singularities, where
we may hope that our approach allows us to
read off symbol alphabets directly from lists of singularity loci.

\subsection{Computational Approaches}

Sec.~2.4 of~\cite{Prlina:2017azl} reviews the Landau equations
and Sec.~6 of that reference details the process of solving them
in several one-loop examples.
Beyond one loop, one approach for seeking
solutions is to perform the analysis ``one loop at a time'', by considering each
one-loop subdiagram and writing down the constraints on the values of other loop and external momenta
imposed by the on-shell and Kirchhoff conditions of the subdiagram.
After taking the union of those constraints, one may conclude that a solution
exists for generic external data,
or that the solution exists only when the
external data satisfy some set of equations.
Solutions of the former type were associated with
the infrared singularities of an amplitude in~\cite{Dennen:2015bet},
and solutions of the latter type indicate branch points of the amplitude
when they live on codimension-one loci in $\Conf_n(\mathbb{P}^3)$.

Here we recall a few basic facts about this loop-by-loop approach,
which has been
carried out for several cases in~\cite{Dennen:2015bet,
Dennen:2016mdk}.

First, as mentioned above, one edge of a bubble subdiagram
can always be removed without affecting
Landau analysis.

Second,
as shown in the Appendix of~\cite{Dennen:2016mdk},
a generic triangle subdiagram
has seven different branches of solutions that should be considered separately.
All of the solutions demand that
the squared sum of momenta on ``external" edges attached to at
least one of the triangle's corners vanish, and the seven
branches of solutions are classified
according to the number of null corners%
\footnote{These corners can be read off as the factors of the Landau
singularity locus, for example in the rightmost column of Tab.~1,
branch (9), of~\cite{Prlina:2017azl}.}.
There are three branches of ``codimension-one" solutions
(any one of the three corners vanishing),
three branches of ``codimension two" (any two of the three corners vanishing)
and one of ``codimension three" (all corners vanishing).
In a Landau diagram analysis, it will often be the case that one of a triangle's corners
is null by fiat; in this case, the solution space will be reduced.
For example, a ``two-mass" triangle subdiagram has only one codimension-one
solution. In the examples we detail in \secRef{two-loop-sample},
all triangle subdiagrams we describe are of this two-mass variety.

Finally, the Kirchhoff conditions associated to a box subdiagram constitute
four homogeneous equations on four Feynman parameters, so the
existence of nontrivial solutions requires the vanishing
of a certain four-by-four determinant called the \emph{Kirchhoff constraint}
for the box.
The Kirchhoff constraints for the four different cases
of box diagrams are summarized in Eqns.~(2.7)
through~(2.11) of~\cite{Dennen:2015bet}.

It is worth noting one detail regarding the ``one loop at a time" approach.
Because the method starts by enumerating the constraints imposed by the
existence of nontrivial solutions to the Landau equations of each subdiagram, it will miss
the solutions which set all Feynman parameters corresponding to some one-loop subdiagram to zero.
However, Landau singularities obtained this way will always be those already present at lower loop
order. So the ``one loop at a time'' approach neglects no novel branch points.
We comment on a specific example of this phenomenon in the next section.

Let us also describe a conceptually simpler but computationally less effective alternative approach
which we have used as a cross-check on our results.
For a given branch of solutions to a set of on-shell conditions, or
equivalently, for a given on-shell diagram, one can reduce the Landau
equations ``all at once'' to see whether they impose codimension-one
constraints on the external data.
This approach is of course usually feasible only with the aid of
a computer algebra system such as Mathematica.
It also lends itself well to numerical experimentation:
one can probe the presence or absence of a putative
singularity at some locus $a=0$ by generating random numeric
values for the external data except for one free parameter $z$,
and then reducing the Landau equations to see if the existence
of nontrivial solutions forces $z$ to take a value that sets
$a = 0$.

Before proceeding to the examples and results,
let us address the question:
how do we confirm that we have detected all singularities?
Starting from the maximal codimension boundaries of the NMHV
amplituhedron
shown in~\tabRef{nmhv-results}, we determine all corresponding Landau diagrams
keeping in mind the ambiguity mentioned in~\secRef{closing}.
From there it is straightforward to produce all possible relaxed Landau diagrams. And from the diagrams we compute the
singularities using the ``one loop at a time'' approach outlined above.
Once we have a list of potential singularities, we turn to the
``all at once'' numerical probing. Doing so we directly confirm
on a diagram-by-diagram basis not only
that the set of singularities is correct, but also that there
are no additional singularities. We have performed these steps
to confirm the NMHV singularities presented in \secRef{symbol-alphabets}.

We will focus only on Landau diagrams that have minimally-NMHV coloring,
as defined in \secRef{on-shell-diagrams-review}, or equivalently,
diagrams that come from a boundary of a two-loop NMHV amplituhedron.
A priori, we cannot dismiss the possibility that a minimally-MHV diagram
may have
novel singularities coming from an NMHV branch of solutions, but
we have explicitly checked that this does not occur in the two-loop NMHV
amplitudes we consider here.
We will demonstrate our ``one loop at a time" approach to solving
Landau equations on an example in the next section, and then proceed to
list the full set of singularities in~\secRef{symbol-alphabets}.

\subsection{A Sample Two-Loop Diagram}
\label{sec:two-loop-sample}

We now turn to the Landau analysis of the boundaries
displayed in~\tabRef{nmhv-results}.
The analysis is very similar to that of the many examples that have been
considered in~\cite{Dennen:2015bet,Dennen:2016mdk}, to which
we refer the reader for additional details.
Therefore we only carry out the analysis in detail
for the case of~\tabRef{nmhv-results}(a), and summarize all
of the results
in the following section.

At maximal codimension
the on-shell conditions
encapsulated in the Landau diagram of~\tabRef{nmhv-results}(a)
are shown in Eqns.~(\ref{eqn:three-mass-color-cut})
and~(\ref{eqn:two-mass-easy-color-cut}).  These have a total of
four discrete solutions, as summarized in~\tabRef{pentagon-box-colorings},
but the only one relevant at NMHV order is the one displayed
in~\eqnRef{nmhvsolution}.
The Landau equations (specifically, the Kirchhoff constraint for
the box subdiagram
defined by~\eqnRef{two-mass-easy-color-cut}) admit a solution
only if~\cite{Dennen:2015bet}
\begin{align}
\langle j(j{-}1\,j{+}1)(i\,i{+}1)\, \mathcal{L}^{(2)}_*\rangle = 0\,.
\end{align}
Substituting in the lower-helicity solution $\mathcal{L}^{(2)}_{*,1}$
and simplifying turns the constraint into
\begin{align}
\langle i\, \overline{j}\rangle
\langle i \, (i{+}1 \, j) \, (k \, k{+}1) \, (k' \, k'{+}1) \rangle = 0\,.
\label{eqn:toreject}
\end{align}

Now we must address a subtlety of the result~(\ref{eqn:toreject})
that is analogous to the one encountered for the maximal
codimension MHV configuration under Eq.~(3.29)
of~\cite{Dennen:2016mdk}.
Like in that case,
the eight-propagator Landau diagram under consideration here,
shown in~\tabRef{nmhv-results}(a), corresponds to a resolution
of a configuration that actually satisfies nine
on-shell conditions, as reviewed in~\secRef{resolutions}.
It was proposed in~\cite{Dennen:2016mdk} that we should trust
the resulting Landau analysis only to the extent that the eight on-shell
conditions imply the ninth for generic external data.
Let us note that if we put
$\langle \mathcal{L}^{(1)}\,\mathcal{L}^{(2)}\rangle=0$
aside for a moment, the NMHV solution to the seven other on-shell
conditions is
\begin{align}
\mathcal{L}^{(2)} = (i\,k\,k{+}1)\cap (i\,k'\,k'{+}1)\,, \qquad
\mathcal{L}^{(1)} = (\alpha Z_i + (1 - \alpha) Z_{i+1}, Z_j)\,,
\end{align}
from which we find
\begin{align}
\langle \mathcal{L}^{(1)}\,\mathcal{L}^{(2)}\rangle
= (1 - \alpha) \langle i(i{+}1\,j)(k\,k{+}1)(k'\,k'{+}1)\rangle\,.
\end{align}
Therefore the conclusion that $\alpha = 1$, and hence
that the ninth condition
$\langle \mathcal{L}^{(1)}\,i{-}1\,i\rangle = 0$
is also satisfied,
actually only follows if
$\langle i(i{+}1\,j)(k\,k{+}1)(k'\,k'{+}1)\rangle \ne 0$.
This observation introduces controversy
about whether the second quantity on the left-hand side of~\eqnRef{toreject} is a valid singularity.
However, note that from the on-shell diagram point of view there is no apparent reason why this singularity
should be excluded, since the diagram can be assigned a valid NMHV coloring
as shown in~\tabRef{pentagon-box-colorings}.
Absent a rigorous argument resolving the matter, we remain agnostic about the status
of this singularity.

It is easy to see that another
solution to the Landau equations with
$\mathcal{L}^{(1)} = (i\,j)$ and $\mathcal{L}^{(2)} = (i\,k\,k{+}1)
\cap (i\,k'\,k'{+}1)$ exists if the four Feynman parameters
associated to the box subdiagram are set to zero.  In this case
the box completely decouples and the pentagon subdiagram reduces to
a three-mass box, so this branch exists if the external data
satisfy the corresponding Kirchhoff constraint
\begin{align}
\label{eqn:threemasskirkhoff}
\langle i(i{-}1\,i{+}1)(k\,k{+}1)(k'\,k'{+}1)\rangle = 0\,.
\end{align}
This illustrates the point highlighted in the previous
section that the ``one loop at a time'' approach can miss certain
solutions to the Landau equations associated entirely with
one-loop subdiagrams.
As mentioned, we are only seeking
new singularities, whereas~\eqnRef{threemasskirkhoff}
is already known from one loop.

\bigskip
\noindent

\begin{figure}
\centering
\begin{tabular}
{
>{\centering\arraybackslash} m{0.35\textwidth}
>{\centering\arraybackslash} m{0.28\textwidth}
>{\centering\arraybackslash} m{0.28\textwidth}
}
(a) & (b) & (c)
\\[-6pt]
\includegraphics{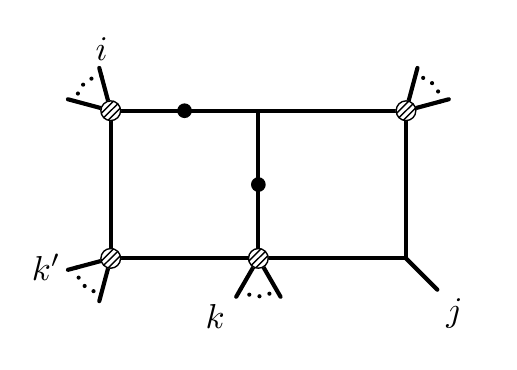}
&
\includegraphics{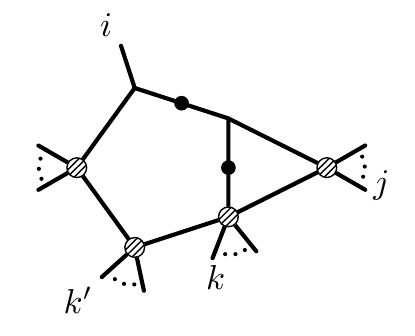}
&
\includegraphics{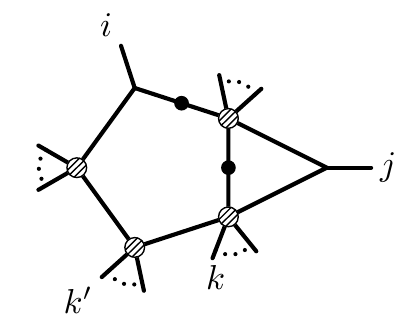}
\\
$\langle \mathcal{L}^{(2)}\,i{-}1\,i\rangle \ne 0$
&
$\langle \mathcal{L}^{(1)}\,j{-}1\,j\rangle \ne 0$
&
$\langle \mathcal{L}^{(1)}\,i\,i{+}1\rangle \ne 0$
\end{tabular}
\caption[Single relaxations]{%
These are the unique single-relaxations of~\tabRef{nmhv-results}(a) that result in $\nkmhv{}$ Landau diagrams.
The computation of the associated Landau singularities is discussed in the text.
}
 \label{fig:single-relaxations}
\end{figure}

Next we move on to codimension seven.
There are four inequivalent relaxations, which we
now discuss in turn.
These relaxations result from collapsing any of the undotted
propagators of~\tabRef{nmhv-results}(a).
We list only the minimally-NMHV diagrams; see~\figRef{single-relaxations}.

\paragraph{Relaxing $\langle \mathcal{L}^{(2)}\,i{-}1\,i\rangle = 0$}
leads to a double-box Landau diagram, \figRef{single-relaxations}(a).

There are two Kirchhoff constraints (one per box), one of which is
easier to determine than the other.
The easier-to-find Kirchhoff constraint comes from the box formed of the $\mathcal{L}^{(1)}$-dependent
propagators (including the shared propagator).  It reads
\begin{equation}
\label{eqn:3akirkhoff1}
\langle j \, (j{-}1 \, j{+}1) \, (i \, i{+}1) \, \mathcal{L}^{(2)}_{*} \rangle = 0 \,,
\end{equation}
where we write $\mathcal{L}^{(2)}_{*}$ to emphasize the loop momentum is
on-shell when all Landau equations are satisfied.

The second Kirchhoff constraint is easiest to find after solving the three $\mathcal{L}^{(1)}$-dependent
on-shell conditions via $\mathcal{L}^{(1)}_{*,1} = (Z_j,B)$, with
$B = \alpha Z_i + (1-\alpha) Z_{i+1}$.
Using this form of $\mathcal{L}^{(1)}$ in the
$\mathcal{L}^{(2)}$-dependent propagators (including the shared one)
results in
\begin{equation}
\langle \mathcal{L}^{(2)} \, i \, B \rangle =
\langle \mathcal{L}^{(2)} \, k \, k{+}1 \rangle =
\langle \mathcal{L}^{(2)} \, k' \, k'{+}1 \rangle =
\langle \mathcal{L}^{(2)} \, j \, B \rangle = 0\,,
\end{equation}
which are now effectively the propagators of a three-mass box.
The second Kirchhoff constraint is therefore
\begin{equation}
\label{eqn:3akirkhoff2}
\langle B \, (i \, j) \, (k \, k{+}1) \, (k' \, k'{+}1) \rangle = 0 \,.
\end{equation}

Solving the remaining on-shell and Kirchhoff constraints
(recall that the three $\mathcal{L}^{(1)}$-dependent conditions were solved already)
fixes
\begin{equation}
\mathcal{L}^{(2)}_{*,1} = (A \, k \, k{+}1) \cap (A \, k' \, k'{+}1) \,, \quad A = (i \, i{+}1) \cap \bar{j} \,, \quad
\textrm{and}
\end{equation}
\begin{equation}
B = (i \, i{+}1) \cap (j \, \mathcal{L}^{(2)}_{*,1}) \,.
\end{equation}
This constraint on $B$ turns~\eqnRef{3akirkhoff2} into a codimension-one constraint on the external data:
\begin{equation}
\langle A \, (i \, j) \, (k \, k{+}1) \, (k' \, k'{+}1)  \rangle = 0 \,, \quad A = (i \, i{+}1) \cap \bar{j} \,,
\end{equation}
which is a new, genuinely two-loop, singularity.

\paragraph{Relaxing $\langle \mathcal{L}^{(1)}\,j{-}1\,j\rangle = 0$}
leads to a pentagon-triangle Landau diagram, \figRef{single-relaxations}(b).

There is a single codimension-one branch for the triangle subdiagram
since there is an on-shell line at one of its corners.
This branch leads to Landau equations with a solution locus
that is a Kirchhoff constraint of three-mass box type:
\begin{align}
\label{eqn:threeboxsingularity}
\langle i \, (i{-}1 \, i{+}1) \, (k\, k{+}1) \, (k'\, k'{+}1) \rangle = 0\,.
\end{align}
We do not focus on these already familiar singularities.

Following any codimension-two branch of the triangle subdiagram leads
to Landau singularities that exist only on
codimension-two loci in the space of external data, which are
not of interest to us.

Following the single codimension-three branch for the triangle leads to
a branch of solutions to the Landau equations that exists only if
\begin{align}
\label{eqn:pentasingularity}
\langle i \, (j\, j{+}1) \, (k\, k{+}1) \, (k'\, k'{+}1) \rangle = 0\,,
\end{align}
which is a new type of singularity.

\paragraph{Relaxing $\langle \mathcal{L}^{(1)}\,i\,i{+}1\rangle = 0$}
leads to a pentagon-triangle Landau diagram, \figRef{single-relaxations}(c).

There is again a single codimension-one branch for the triangle subdiagram
leading to an effective decoupling of the two loop momenta
and an overall Landau constraint of the same
form (up to relabeling) as~\eqnRef{threeboxsingularity}.

Following the codimension-two branches for the triangle subdiagram
uncovers constraints of codimension higher than one
on the external data, which cannot sensibly be associated with
branch points.

Following the codimension-three branch for the triangle subdiagram leads to the
same Landau singularity as in~\eqnRef{pentasingularity} (up to relabeling).

%
%
%
%
%
%

\begin{figure}
\centering
\begin{tabular}
{
>{\centering\arraybackslash} m{0.32\textwidth}
>{\centering\arraybackslash} m{0.32\textwidth}
>{\centering\arraybackslash} m{0.225\textwidth}
}
\\
(a) & (b) & (c)
\\[-6pt]
\includegraphics{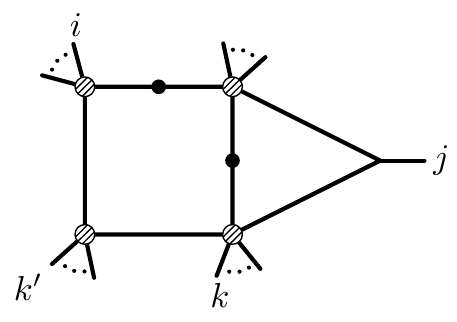}
&
\includegraphics{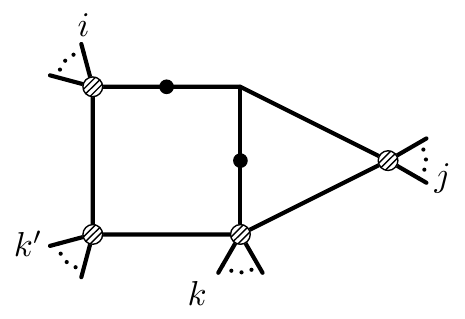}
&
\includegraphics{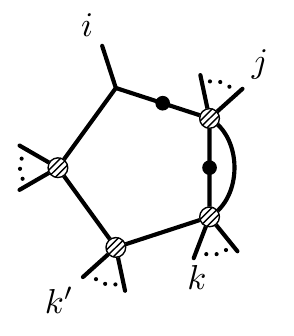}
\\
$\langle \mathcal{L}^{(2)}\,i{-}1\,i\rangle \ne 0$
$\langle \mathcal{L}^{(1)}\,i\,i{+}1\rangle \ne 0$
&
$\langle \mathcal{L}^{(2)} \, i{-}1 \, i \rangle \ne 0$
$ \langle \mathcal{L}^{(1)} \, j{-}1 \, j \rangle \ne 0$
&
$\langle \mathcal{L}^{(1)}\,i\,i{+}1\rangle \ne 0$
$\langle \mathcal{L}^{(1)}\,j{-}1\,j\rangle \ne 0$
\\
\end{tabular}
\caption[Double relaxations]{%
These are the unique minimally NMHV relaxations of
the diagrams~\figRef{single-relaxations}.
As such, these are also
double-relaxations of~\tabRef{nmhv-results}(a).
Computing the associated singularities is discussed in the text.
Any further relaxations of triangles yield bubble subdiagrams.
In (c), relaxing either of $\langle \mathcal{L}^{(2)}\,i\,i{\pm}1\rangle= 0$ yields the four-mass bubble-box of~\figRef{bubble-box}.
}
 \label{fig:double-relaxations}
\end{figure}

\bigskip
\noindent
At codimension six there are three inequivalent relaxations,
shown in~\figRef{double-relaxations},
that do not reduce the Landau diagram to an MHV one.
Collapsing any of the undotted propagators of a box subdiagram in~\figRef{double-relaxations}
results in a minimally-MHV Landau diagram, as one of the external labels would
necessarily drop out.
Any additional relaxations of a propagator
in a triangle subdiagram of~\figRef{double-relaxations} will yield a bubble
subdiagram,
which cannot yield a new singularity as we have already emphasized.

\paragraph{Relaxing both $\langle \mathcal{L}^{(1)} \,i \, i{+}1  \rangle$ = $\langle \mathcal{L}^{(2)} \, i{-}1 \, i \rangle = 0$}
leads to a box-triangle Landau diagram, \figRef{double-relaxations}(a).

The single codimension-one branch of the triangle
leads to the effective decoupling of the two loops
and results in Landau singularities at Mandelstam-type loci:
\begin{align}
\langle i \, i{+}1 \, k \, k{+}1 \rangle \langle i \, i{+}1 \, k' \, k'{+}1 \rangle \langle k \, k{+}1 \, k' \, k'{+}1 \rangle = 0\,.
\end{align}

The same Landau singularities are obtained by following the codimension-two branches for the triangle.

Following the codimension-three branch for the triangle leads to the constraint
\begin{align}
\langle j \, (i \, i{+}1) \, (k \, k{+}1) \, (k' \, k'{+}1) \rangle = 0\,.
\end{align}

\paragraph{Relaxing both $\langle \mathcal{L}^{(2)} \, i{-}1 \, i \rangle = \langle \mathcal{L}^{(1)} \, j{-}1 \, j \rangle = 0$}
leads to a box-triangle Landau diagram, \figRef{double-relaxations}(b).
All branches of the triangle subdiagram
result in bubble-type singularities, $\langle a \, a{+}1 \, b \, b{+}1 \rangle$, or
higher codimension constraints.

\paragraph{Relaxing both $\langle \mathcal{L}^{(1)} \, i \, i{+}1 \rangle = \langle \mathcal{L}^{(1)} \, j{-}1 \, j \rangle = 0$}
leads to a pentagon-bubble Landau diagram, \figRef{double-relaxations}(c),
as discussed above and displayed in~\figRef{bubble-box}, with a singularity
on the locus
\begin{align}
\langle i \, (j \, j{+}1) \, (k \, k{+}1) \, (k' \, k'{+}1) \rangle = 0\,.
\end{align}

\begin{figure}
\centering
\begin{tabular}
{
>{\centering\arraybackslash} m{0.3\textwidth}
>{\centering\arraybackslash} m{0.3\textwidth}
>{\centering\arraybackslash} m{0.3\textwidth}
}
\\
(a) & (b) & (c)
\\[-6pt]
\includegraphics{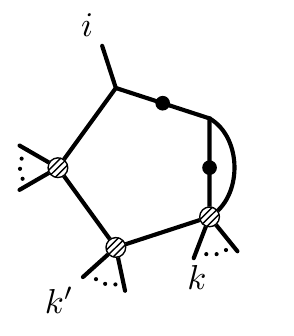}
&
\includegraphics{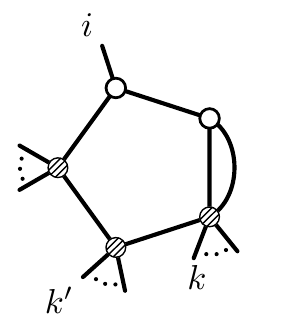}
&
\includegraphics{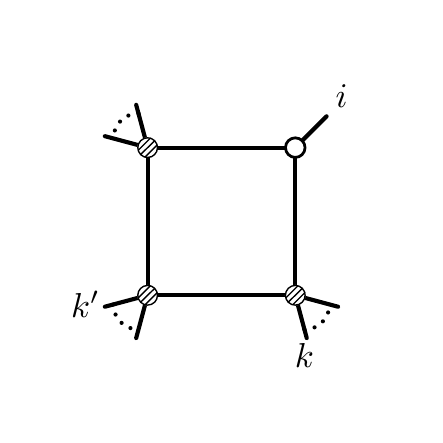}
\\
$\langle \mathcal{L}^{(1)}\,j{-}1\,j\rangle \ne 0$
$\langle \mathcal{L}^{(1)}\,j\,j{+}1\rangle \ne 0$
&
Minimal Coloring
&
After Graph Moves
\\
\end{tabular}
\caption[Last NMHV double relaxation.]{%
The Landau diagram (a) appears to have a minimally MHV coloring (b).
Yet the corresponding on-shell function is related by the on-shell
diagram moves of \cite{ArkaniHamed:2012nw} to one in the NMHV helicity sector (c).
}
 \label{fig:last-double-relaxation}
\end{figure}

\paragraph{Relaxing both $\langle \mathcal{L}^{(1)} \, j{-1} \, j \rangle = \langle \mathcal{L}^{(1)} \, j \, j{+}1 \rangle = 0$}
is displayed in \figRef{last-double-relaxation}.
This case is interesting because it emphasizes the interplay between on-shell diagrams and the amplituhedron.

From the on-shell diagram perspective, this diagram naively has a minimally MHV coloring,
\figRef{last-double-relaxation}(b).
However the graph moves that preserve on-shell functions
(particularly the ``collapse and re-expand'' and ``bubble deletion'' of Sec.~2.6 of \cite{ArkaniHamed:2012nw})
permit redrawing the coloring as a three-mass box on-shell diagram  \figRef{last-double-relaxation}(c),
colored in its minimal helicity manner, $\hel \ge 1$ .
Since the graph moves preserve the on-shell function,
the original on-shell diagram must also be minimally NMHV.

It is straightforward to check that the momentum twistor solution
corresponding to this minimal coloring \figRef{last-double-relaxation}(a)
is in fact a boundary of an NMHV amplituhedron, not an MHV one,
and so the on-shell diagram and amplituhedron perspectives align.

For the two-loop amplitude, this diagram does not contribute new
possible branch points, but this phenomenon is something to keep
in mind for future studies.

\bigskip
\noindent
There are no new NMHV triple relaxations,
but we revisit a case discussed earlier to show how it naturally
arises in this organizational scheme.

\paragraph{Relaxing all of
$\langle \mathcal{L}^{(2)} \, i{-}1 \, i \rangle =
\langle \mathcal{L}^{(1)} \, i \, i{+}1 \rangle =
\langle \mathcal{L}^{(1)} \, j{-}1 \, j \rangle = 0$}
leads to the bubble-box Landau diagram discussed above and displayed in~\figRef{bubble-box}.
As mentioned above, this does not contribute a new two-loop singularity
but it does indicate that two-loop NMHV amplitudes inherit
the four-mass box singularity that appears at one loop only
starting at $\hel=2$.
Our analysis indicates this is a fairly common phenomenon:
Landau diagrams for an $L$-loop $\nkmhv{\hel}$ amplitude
that contain bubble or triangle subdiagrams will often
contain singularities that also contribute
to $(L-1)$-loop $\nkmhv{\hel+1}$ amplitudes.

\subsection{Two-Loop NMHV Symbol Alphabets}
\label{sec:symbol-alphabets}

The full set of loci in the external kinematic space
$\Conf_n(\mathbb{P}^3)$ where two-loop NMHV amplitudes have
Landau singularities is obtained by carrying out the analysis
of the previous section for all Landau diagrams appearing
in Tabs.~\ref{tab:mhv-results} and~\ref{tab:nmhv-results},
together with all of their (still NMHV) relaxations.
Among the set of singularities generated in this way are the two-loop MHV singularities
that arise from the configuration shown in~\tabRef{mhv-results},
which live on the loci
\begin{align} \begin{split}
\langle a \, a{+}1\, b\, c \rangle = 0\,, \\
\langle a \, a{+}1\, \overline{b} \cap \overline{c} \rangle = 0\,,
\label{eqn:twoloopalphabet}
\end{split} \end{align}
for arbitrary indices $a, b, c$.
The set of brackets appearing on the left-hand sides
of~\eqnRef{twoloopalphabet} correspond exactly to the set
of symbol letters of two-loop MHV amplitudes originally
found in~\cite{CaronHuot:2011ky}.

For the NMHV configurations shown in~\tabRef{nmhv-results}
we find additional singularities that live on loci of the form\footnote{Out
of caution we have included on the first line the singularities of
the type shown in~\eqnRef{toreject}; but we remind the reader of the discussion in
the subsequent
paragraph; for $n=8$ it happens that the first line is necessarily
a particular case of the second and/or fourth so there is no controversy.}
\begin{align}
\begin{split}
\langle i\,(i{\pm}1\,\ell)(j\,j{+}1)(k\,k{+}1)\rangle &= 0\,,\\
			\langle j \, (j{-}1\, j{+}1) \, (j'\, j'{+}1) \, (i\, \ell) \rangle & =0 \,, \\
			\langle i\,(j\,j{+}1)(k\,k{+}1)(\ell\,\ell{+}1)\rangle & = 0 \,, \\
			\langle i\,i{+}1\, \overline{j} \cap (k \, k' \, k'{+}1)     \rangle &= 0 \,, \\
			\langle \overline{i} \cap (i\,i'\, i'{+}1) \, \cap \, \overline{j} \cap (j\,j'\,j'{+}1) \rangle &= 0 \,, \\
          \llangle (i\,i{+}1) \cap \overline{j};(i\,j)(k\,k{+}1)(\ell\,\ell{+}1)\rrangle & = 0  \,,
\end{split}
\label{eqn:nmhvsymbolalphabets}
\end{align}
using notation explained in Appendix~\ref{sec:notation}.
The indices are restricted (as a consequence of planarity)
to have the cyclic
ordering $\ell \le  \{ i, i'\} \le \{ j, j' \} \le \{k, k'\} \le \ell$
(or the reflection of this, with all $\le$'s replaced by $\ge$'s)
where the curly bracket notation means that the relative ordering of an index
with its primed partner is not fixed (tracing back to the
ambiguity discussed in~\secRef{closing}).

In addition to singularities of the type listed
in~\eqnRef{nmhvsymbolalphabets}, two-loop NMHV amplitudes
also have four-mass box singularities as discussed in the
beginning of~\secRef{nmhv-landau-analysis} and illustrated
in~\figRef{bubble-box}.
Although guessing symbol letters from knowledge of singularity
loci is in general nontrivial (see Sec.~7
of~\cite{Prlina:2017azl}),
we conjecture that
the quantities appearing on the left-hand sides
of Eqns.~(\ref{eqn:twoloopalphabet}) and~(\ref{eqn:nmhvsymbolalphabets}),
together with appropriate symbol letters of four-mass box type (see
the example in the following section),
constitute the symbol alphabet of two-loop NMHV amplitudes in
SYM theory.
It is to be understood that all degenerations
of the indicated forms are meant to be included as well, for example
such as taking $j=j'-1$ in the first line.  For certain values of
some indices the expressions can degenerate into symbol letters
(or products of symbol letters) that already
appear in~\eqnRef{twoloopalphabet}, or
elsewhere in~\eqnRef{nmhvsymbolalphabets}, but other degenerate cases are
valid, new NMHV letters.

It is interesting to note that for arbitrary $n$ the conjectural set of
symbol letters in~\eqnRef{nmhvsymbolalphabets}
is not closed under parity, unlike the
two in~\eqnRef{twoloopalphabet} which are parity conjugates of each
other\footnote{More precisely, the parity conjugate of the first
quantity in~\eqnRef{twoloopalphabet}
is $\langle a{-}1\,a\,a{+}1\,a{+}2\rangle$ times the second;
they become exactly parity conjugate in a gauge where the
momentum twistors are scaled so that all four-brackets of
four adjacent indices are set to 1.}.
We know of no a priori reason why the symbol alphabet for a given
amplitude in SYM theory should be closed under parity;
in principle,
the parity symmetry of the theory requires
only that the symbol alphabet of $\nkmhv{\hel}$
amplitudes must be the parity conjugate of the symbol alphabet
of $\nkmhv{n-\hel-4}$ amplitudes.

The absence of parity symmetry is a simple consequence of the fact that
different branches of solutions to the Landau equations give non-zero
support to amplitudes in different helicity sectors (or,
equivalently, overlap boundaries of amplituhedra in different helicity sectors).
From this point of view it appears to be an accident that the
two-loop MHV symbol alphabet is closed under parity; we guess
that this will continue to hold at arbitrary loop order.
It is also an interesting consistency check that for $n < 8$ the symbol
letters in~\eqnRef{nmhvsymbolalphabets} necessarily degenerate
into letters of the type already present at MHV order.
This is consistent with all results available to date from
the hexagon and heptagon amplitude bootstrap programs, which are
based on the hypothesis that the symbol alphabet for all amplitudes
with $n < 8$ is given by~\eqnRef{twoloopalphabet} to all loop order.
Genuinely new NMHV letters begin to appear only starting at $n=8$,
to which we now turn our attention.

\subsection{Eight-point Example}

For the sake of illustration let us conclude by explicitly
enumerating our conjecture for the two-loop NMHV symbol
alphabet for the case $n=8$.  First let us recall that
the corresponding MHV symbol alphabet~\cite{CaronHuot:2011ky}
is comprised of 116 letters:
\begin{itemize}
\item 68 four-brackets of the form
$\langle a\, a{+}1\, b\, c\rangle$ (there are altogether
of $\binom{8}{4}=70$ four-brackets
of the more general form
$\langle a\, b\, c\, d\rangle$, but at $n=8$
both
$\langle 1\, 3\, 5\, 7\rangle$ and
$\langle 2\, 4\, 6\, 8\rangle$ are excluded by the requirement
that at least one pair of indices must be adjacent),
\item 8 cyclic images
of $\langle 1\,2\,\overline{4}  \cap \overline{6}\rangle$,
\item
and 40 degenerate cases of $\langle a\, a{+}1\, \overline{b} \cap
\overline{c}\rangle$ consisting of
8 cyclic images each of
$\langle 1\,(2\,3)(4\,5)(7\,8)\rangle$,
$\langle 1\,(2\,3)(5\,6)(7\,8)\rangle$,
$\langle 1\,(2\,8)(3\,4)(5\,6)\rangle$,
$\langle 1\,(2\,8)(3\,4)(6\,7)\rangle$,
as well as
$\langle 1\,(2\,8)(4\,5)(6\,7)\rangle$.
\end{itemize}
Referring the reader again to Appendix~\ref{sec:notation} for
details on our notation,
we conjecture that an additional 88 letters appear
in the symbol alphabet of the two-loop $n=8$ NMHV amplitude\footnote{The
$116+88=204$ symbol letters of this amplitude can
be assembled into
$204-8=196$ dual conformally invariant
cross-ratios in many different ways.  We
cannot \emph{a priori} rule out the possibility that the symbol
of this amplitude might be expressible in terms of an even smaller
set of carefully chosen multiplicatively independent
cross-ratios, though this type of reduction is not possible
in any known six- or seven-point examples.}
\begin{itemize}
\item 48 degenerate cases consisting of 16 dihedral images
each of
$\langle 1\,(2\,3)(4\,5)(6\,7) \rangle$,
$\langle 1\,(2\,3)(4\,5)(6\,8) \rangle$,
as well as
$\langle 1\,(2\,8)(3\,4)(5\,7) \rangle$,
\item
8 cyclic images of
$\langle \overline{2} \cap (2\,4\,5) \cap \overline{8} \cap (8\,5\,6)\rangle$
(this set is closed under reflections, so adding all dihedral
images would be overcounting),
\item the 8 distinct dihedral images of
$\langle \overline{2} \cap (2\,4\,5) \cap \overline{6} \cap (6\,8\,1)\rangle$
(which is distinct from its reflection but comes back to itself
after cycling the indices by four),
\item 16 dihedral images of
$\llangle (1\,2)\cap\overline{4};(1\,4)(5\,6)(7\,8)\rrangle$,
\item and finally 8
four-mass box-type letters.
\end{itemize}
The last of these were displayed in Eq.~(7.1)
of~\cite{Prlina:2017azl} and take the form
\begin{align}
f_{i\ell}f_{jk} \pm( f_{ik} f_{j\ell}- f_{ij}
f_{k\ell}) \pm \sqrt{(f_{ij} f_{k\ell} - f_{ik} f_{j\ell} + f_{i\ell} f_{jk})^2
-4 f_{ij} f_{jk} f_{k\ell} f_{i\ell}}\,,
\end{align}
where $f_{ij} \equiv \langle i\, i{+}1\, j\, j{+}1\rangle$
and the signs may be chosen independently.
For $n=8$ there are two inequivalent
choices $\{i,j,k,\ell\} = \{1,3,5,7\}$ or
$\{2,4,6,8\}$, for a total of eight possible symbol letters of this type.

\section{Conclusion}

The symbol alphabets for all two-loop MHV amplitudes in SYM theory
were first found in~\cite{CaronHuot:2011ky}.
In~\cite{Dixon:2011nj,CaronHuot:2011kk}
it was found that two-loop NMHV amplitudes have the same
symbol alphabets as the corresponding MHV amplitudes
for $n=6, 7$, which is now believed to be true to all loop order.
However, the question of whether two-loop NMHV amplitudes
for $n>7$ have the same symbol alphabets as their MHV
cousins has remained open.
In this paper we find that the former have branch points
(of the type shown in~\eqnRef{nmhvsymbolalphabets})
not shared by the latter, answering this question in the negative.

Our conjectures for the two-loop NMHV symbol alphabets
are formulated
in terms of quantities analogous to the
cluster $\mathcal{A}$-coordinates of~\cite{Golden:2013xva},
although it is simple to confirm that at least some of them are not
cluster coordinates of the $\Gr(4,n)$ cluster algebra (it is possible
that none of them are, but some of them are more difficult to check).
For the purpose of carrying out the amplitude bootstrap, it is
however more convenient to assemble these letters into
dual conformally invariant cross-ratios.
In the literature considerable effort (see
for example~\cite{Golden:2013lha,Golden:2014xqa,Golden:2014pua,Harrington:2015bdt,Drummond:2017ssj})
has gone into divining deep mathematical structure of
amplitudes hidden in the particular kinds of cross-ratios
that might appear, especially when they can be taken to
cluster $\mathcal{X}$-coordinates (or Fock-Goncharov coordinates)
of the type reviewed in~\cite{Golden:2013xva}.
However, we see no hint in the Landau analysis
or inherent to the twistor or on-shell diagrams employed
in this paper that suggests any preferred way of
building such cross-ratios.

It is inherent in the approach taken here
following~\cite{Dennen:2016mdk,Prlina:2017azl} (as well as in the
amplitude bootstrap program itself) that we eschew
knowledge of or interest in explicit representations of
amplitudes in terms of local Feynman integrals.
However, as mentioned in the conclusion of~\cite{Prlina:2017azl},
the procedure of identifying relevant boundaries of amplituhedra
and then solving the Landau equations associated to each one
as if it literally represented some Feynman integral
is suggestive that this approach might be thought of as naturally
generating integrand expansions
around the highest codimension amplituhedron boundaries\footnote{We are
grateful to N.~Arkani-Hamed for extensive discussions on this point.}.
This approach might lead to a resolution of the controversy
regarding the status of Landau singularities
of the type~\eqnRef{toreject}
obtained from maximal codimension boundaries.
This analysis is, however, beyond the scope
of our paper and we remain agnostic about the status of this branch point in anticipation of
empirical data.
If this singularity is shown to be spurious, this would be an interesting result not easily explainable using on-shell diagram techniques, and it would signal that boundaries of amplituhedra contain more information waiting to be explored.

These observations highlight a point that we
have emphasized several times in this
paper and the prequel~\cite{Prlina:2017azl}.
Namely, several threads in this tapestry, including
the connection to on-shell diagrams
reviewed in~\secRef{on-shell-diagrams}
and the simple relation between twistor diagrams
and Landau diagrams in Appendix~\ref{sec:twistor-to-landau},
do not inherently rely on planarity.
This hints at the tantalizing possibility that
some of our toolbox may be useful for studying non-planar
amplitudes about which much less is known
(see~\cite{Arkani-Hamed:2014via,Bern:2014kca,Bern:2015ple}).

One of the stronger hints
--- the relationship between on-shell diagrams and Landau diagrams ---
also aids in corroborating results.
A vanishing on-shell diagram indicates a
location where the analytic structure of an amplitude is trivial;
that is exactly the same information encoded by the boundaries of the amplituhedron.
The simple connection between the results tabulated in~\secRef{presentation}
and those obtained via the
on-shell diagram approach provides an important cross-check
supporting the validity of our analysis, as well as giving
additional corroboration to the definition of amplituhedra.

\acknowledgments

We have benefited greatly from
very stimulating discussions with N.~Arkani-Hamed, L.~Dixon
and J.~Bourjaily, and from
collaboration with A.~Volovich in the early stages of this work.
This work was supported in part by: the US Department of Energy under
contract DE-SC0010010 Task A,
Simons Investigator Award \#376208 of
A.~Volovich (JS),
the Simons Fellowship Program in Theoretical Physics (MS),
the National Science Foundation under Grant No. NSF PHY-1125915 (JS),
and the Munich Institute for Astro- and Particle
Physics (MIAPP) of the DFG cluster of excellence ``Origin and Structure
of the Universe'' (JS).
MS is also grateful to the CERN theory group for hospitality
and support during the course of this work.

\appendix

\section{Notation}
\label{sec:notation}

Here we recall some standard momentum twistor notation
and define some new notation used in~\secRef{nmhv-landau-analysis}.
The momentum twistors $Z_a^I$ are $n$ homogeneous
coordinates on $\Conf_n(\mathbb{P}^3)$ (so
$I \in \{1,\ldots,4\}$ and
$a \in \{1,\ldots,n\}$) in terms of which we have
the natural four-brackets
\begin{align}
  \langle a\, b\, c\, d \rangle \equiv \epsilon_{IJKL}
  Z_a^I Z_b^J Z_c^K Z_d^L\,.
\end{align}
We use (see for example Eq.~(2.38) of~\cite{ArkaniHamed:2010gh})
\begin{align}
\langle x\,y\,(a\,b\,c)\cap(d\,e\,f)\rangle
\equiv \langle x\,a\,b\,c\rangle \langle y\,d\,e\,f\rangle -
\langle y\,a\,b\,c\rangle \langle x\,d\,e\,f\rangle
\end{align}
and in the special case when the two planes
$(a\,b\,c)$, $(d\,e\,f)$
share a common
point, say $f = c$, we use the shorthand
\begin{align}
\label{eqn:degenerate}
\langle c\, (x\,y) (a\,b) (d\,e)\rangle \equiv
\langle x\,y\, (a\,b\,c)\cap (d\,e\,c)\rangle
\end{align}
to emphasize the otherwise non-manifest fact that this
quantity is fully antisymmetric under the exchange
of any two of the three lines $(x\,y)$, $(a\,b)$, and $(d\,e)$.
In~\secRef{nmhv-landau-analysis} we introduce a bracket
for the intersection of four planes which is related by
the obvious duality to an intersection of four points.
Specifically, if we represent a plane $(a\,b\,c)$ by its dual point
\begin{align}
(a\,b\,c)_I \equiv \epsilon_{IJLK} Z_a^J Z_b^K Z_c^L
\end{align}
then we define
\begin{multline}
\langle (a_1\,a_2\,a_3)\cap
(b_1\,b_2\,b_3)\cap
(c_1\,c_2\,c_3)\cap
(d_1\,d_2\,d_3)\rangle
\\
\equiv \epsilon^{IJKL}
(a_1\,a_2\,a_3)_I
(b_1\,b_2\,b_3)_J
(c_1\,c_2\,c_3)_K
(d_1\,d_2\,d_3)_L\,.
\end{multline}
Our final new definition
\begin{align}
\label{eqn:weird}
\llangle (a\,a{+}1) \cap \overline{b};
(a\,b)(c\,d)(e\,f)\rrangle
\equiv \frac{\langle ((a\,a{+}1) \cap \overline{b}) (a\,b)
(c\,d)(e\,f) \rangle}{\langle a\,\overline{b}\rangle}
\end{align}
requires a little bit of explanation.  The first quantity in the
numerator recalls that the intersection of a line
$(a\,b)$ and a plane $(c\,d\,e)$ can be represented by the point
(see for example p.~33 of~\cite{ArkaniHamed:2010gh})
\begin{align}
(a\,b) \cap (c\,d\,e) \equiv Z_a \langle b\,c\,d\,e \rangle
+ Z_b \langle c\,d\,e\,a\rangle\,.
\end{align}
Using this definition, the $(a\,a{+}1)\cap \overline{b}$
in the numerator of~\eqnRef{weird} defines a point
that feeds into ``$c$'' in the definition~(\ref{eqn:degenerate}).
By following the trail of definitions
it is easy to check that the resulting bracket
in the numerator of~\eqnRef{weird} always has an overall
factor of $\langle a\,\overline{b}\rangle$, which we divide
out in order to make $\llangle . \rrangle$ irreducible (for
general arguments).

\section{Twistor Diagrams to Landau Diagrams}
\label{sec:twistor-to-landau}

%
%
%
\begin{figure}
\centering
\begin{tabular}
{
>{\centering\arraybackslash} m{0.27\textwidth} 
>{\centering\arraybackslash} m{0.315\textwidth} 
>{\centering\arraybackslash} m{0.315\textwidth} 
}
\includegraphics{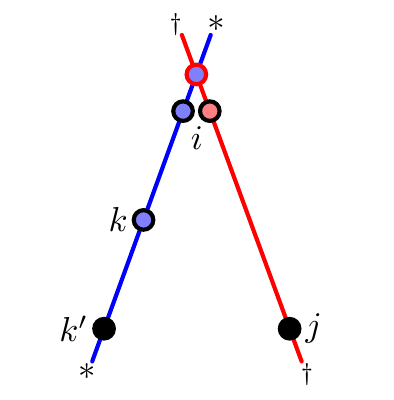}
&
\includegraphics{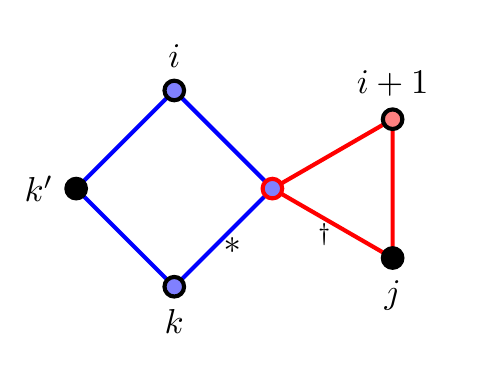}
&
\includegraphics{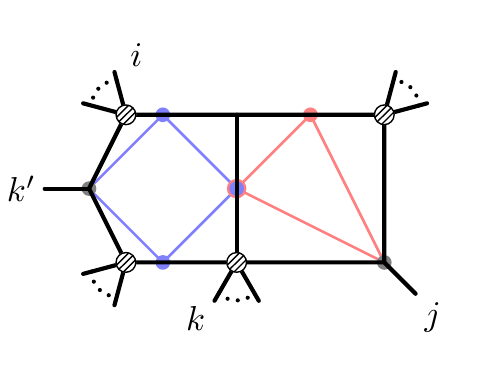}
\\
(a) & (b) & (c)
\end{tabular}
\caption[Edge-node duality]{%
Figure (a) is momentum twistor diagram (with
the superfluous edges associated with the external labels
dropped), with the endpoints of {\color{red} $\mathcal{L}^{(1)}$} (denoted with $\dagger$)
identified, and with  the endpoints of {\color{blue} $\mathcal{L}^{(2)}$} (denoted with *)
identified. The resulting graph (b) comes from preserving the ordering of nodes along the line.
Note we have used the ambiguity in ordering between $k$ and $k'$ to consider the case
$k < k' < i$.
Mapping any empty node to an edge between massive corners
and mapping any filled node to a massless corner with two edges between massive corners,
results in the Landau diagram (c).
There would be a massive corner along the top of the Landau diagram,
but the indices fix that corner to be zero.
}
 \label{fig:edge-node-duality}
\end{figure}
%
%
%

In this appendix, we explain how twistor diagrams and Landau diagrams
are partial edge-to-node duals of each other.
This map straightforwardly generalizes to any number of loops,
and is easily inverted to map a Landau diagram into a twistor diagram.

We first note that the edges in a twistor diagram associated with the external labels are redundant,
since the ``empty'' or ``filled'' property of the node already tracks the same information.
So we can drop such edges.
Then the following steps map a twistor diagram, $\tau$, to a Landau diagram, $\lambda$:
\begin{enumerate}
\item For each loop line $\mathcal{L}^{(i)}$ in $\tau$, identify one endpoint of $\mathcal{L}^{(i)}$
with the other endpoint of $\mathcal{L}^{(i)}$. Since $\tau$ are graphs,
this identification preserves the order of the nodes along all $\mathcal{L}^{(i)}$.
\item Map each empty $\tau$-node into a $\lambda$-edge.
Identify the two $\lambda$-nodes defining the $\lambda$-edge as massive corners of $\lambda$.
\item Map each filled $\tau$-node into two $\lambda$-edges sharing one common $\lambda$-node.
Identify the common $\lambda$-node as a massless corner of $\lambda$,
and the other two $\lambda$-nodes as massive corners of $\lambda$.
\item The external labels map from $\tau$ to $\lambda$ such that:
\begin{itemize}
\item the label of an empty $\tau$-node maps to one of the two massless corners defining the new $\lambda$-edge, and
\item the label of a filled $\tau$-node maps into a massless corner of $\lambda$.
\end{itemize}
\end{enumerate}
This is a partial edge-to-node map because only
the empty $\tau$-nodes obey a proper edge-to-node exchange as they map to a $\lambda$-edge,
while the filled $\tau$-nodes are effectively unchanged as they map to $\lambda$-nodes.
It is always possible to consistently assign the labels of $\tau$ to $\lambda$,
though so doing may cause a massive corner of $\lambda$ to completely vanish,
as happens in the following concrete example.

We turn now to detailing how the two-loop twistor diagram, \figRef{edge-node-duality}(a),
(also the first column of~\tabRef{nmhv-results}(b))
maps into one of its corresponding Landau diagrams, \figRef{edge-node-duality}(c)
(the second column of~\tabRef{nmhv-results}(b)).

The first step is to identify the two nodes
corresponding to the end-points of each $\mathcal{L}^{(i)}$, $i=1,2$.
This closes the two lines into loops, and we formally think of the diagram
as a graph, specified by its edges, nodes, and decorations of its nodes.
The result is~\figRef{edge-node-duality}(b), with identified endpoints marked by * and $\dagger$.

In this instance, there is an ambiguity in choosing $k'<k$ or $k<k'$.
We demonstrate the latter case here to highlight that $k'$ and $k$ can be swapped
with respect to how they appear on the loop line.
The $k'<k$ differs from what we detail here by swapping ordering of the two nodes along the loop.
Then the filled $k'$ node would be on the bottom of the box in~\figRef{edge-node-duality}(b),
while the empty $k$ node would be on the left of the box.

In the $k<k'$ case we are considering,
the resulting graph becomes the Landau diagram, \figRef{edge-node-duality}(c)
under the partial edge-to-node dual map,
as described in the steps above. Note in~\figRef{edge-node-duality}(c) that
the empty nodes of the original twistor diagram are identified with edges of the resulting
Landau diagram. In contrast, the filled nodes are identified with massless corners,
which are themselves nodes. So this is only a partial edge-to-node map.

\end{document}